\documentclass[12pt]{article} 
\usepackage{graphicx}
\newcommand{\be}{\begin{equation}}
\newcommand{\ee}{\end{equation}}
\newcommand{\bear}{\begin{eqnarray}}
\newcommand{\eear}{\end{eqnarray}}
\newcommand{\ba}{\begin{array}}
\newcommand{\ea}{\end{array}}

\newcommand{\CN}{{\cal N}}

\begin{document}

\begin{titlepage}
\vfill
\begin{flushright}
{\normalsize IC/2009/016}\\
{\normalsize arXiv:0903.4894 [hep-th]}\\
\end{flushright}

\vfill
\begin{center}
{\Large\bf Holographic nonlinear hydrodynamics from AdS/CFT with multiple/non-Abelian symmetries  }

\vskip 0.3in

{ Mahdi Torabian\footnote{\tt mahdi@ictp.it}, Ho-Ung
Yee\footnote{\tt hyee@ictp.it}}

\vskip 0.15in

 {\it ICTP, High Energy, Cosmology and Astroparticle Physics,} \\
{\it Strada Costiera 11, 34014, Trieste, Italy}
\\[0.3in]

{\normalsize  2009}

\end{center}

\vfill

\begin{abstract}

We study viscous hydrodynamics of hot conformal field theory plasma with multiple/non-Abelian symmetries
in the framework of AdS/CFT correspondence, using a recently proposed method
of directly solving bulk gravity in derivative expansion of local plasma parameters.
Our motivation is to better describe the real QCD plasma produced at RHIC, incorporating its $U(1)^{N_f}$
flavor symmetry as well as $SU(2)_I$ non-Abelian iso-spin symmetry.
As concrete examples, we choose to study the STU model for multiple $U(1)^3$ symmetries, which is
a sub-sector of 5D N=4 gauged SUGRA dual to N=4 Super Yang-Mills theory, capturing Cartan $U(1)^3$ dynamics inside
the full R-symmetry.
For $SU(2)$, we analyze the minimal 4D N=3 gauged SUGRA whose bosonic action is simply
an Einstein-Yang-Mills system, which corresponds to $SU(2)$ R-symmetry dynamics on M2-branes at a Hyper-Kahler cone.
By generalizing the bosonic action to arbitrary dimensions and Lie groups, we present our analysis and results for
any non-Abelian plasma in arbitrary dimensions.

\end{abstract}

\vfill

\end{titlepage}
\setcounter{footnote}{0}

\baselineskip 18pt \pagebreak
\renewcommand{\thepage}{\arabic{page}}
\pagebreak

\section{Introduction and summary}

Strongly coupled plasma of finite temperature gauge theories has
recently become a fascinating subject of research, largely motivated
by the RHIC experiment of relativistic heavy ion collisions. Naive
QCD expectations based on perturbative QCD have failed to explain
certain important aspects of the created QCD plasma, and there are
several indications that the RHIC plasma is in fact a strongly
coupled liquid. Given the situation, one may hope that the problem
can be attacked by AdS/CFT correspondence or gauge/gravity
correspondence because the correspondence is useful precisely when
the gauge theory side is strongly coupled \cite{Shuryak:2005ia}. In the gravity side, a
finite temperature plasma corresponds to a black-hole, or more
precisely black-brane, spacetime with Hawking temperature identified
with the temperature of the gauge theory. The black-hole horizon is
located at certain point in the holographic additional dimension,
and presumably physics outside the horizon with suitable boundary
conditions on the horizon describes the finite temperature plasma of
gauge theories. There have been a lot of useful and often surprising
results obtained from this gravity picture, which would be hard to
be found in the pure gauge theory analysis
due to strong coupling \cite{Policastro:2001yc,Kovtun:2003wp,Buchel:2003tz,Janik:2005zt,Heller:2008mb,Hartnoll:2007ip}.

Of course, the main huddle that hinders further progress in this
direction is the absence of precise dual gravity theory of the real
QCD at present, although it is important to improve the current
models to better mimic realistic QCD \cite{Sakai:2004cn,Erlich:2005qh,Kiritsis:2009hu,Torabian:2008dh}. However, certain properties of
strongly coupled finite temperature plasma may be universal at least
qualitatively \cite{Buchel:2003tz}; a well-known example is the
viscosity-entropy ratio \cite{Policastro:2001yc,Kovtun:2003wp,Buchel:2003tz},
$\eta/s\sim {1\over 4\pi}$, which in fact is close to the RHIC
experiment data, and it is not a vividly wrong idea to try to learn
something about realistic RHIC plasma by studying certain specific
AdS/CFT models at finite temperature\footnote{See \cite{Buchel:2008ae} for possible violations of viscosity-entropy bound.}.
The question whether the
results obtained in the specific model are meaningful in realistic
QCD should be asked carefully though.

Any finite temperature plasma is described by hydrodynamics in
sufficiently slowly-varying and long-ranged regime. It is more a
framework rather than a result; it is based on local thermal
equilibrium and conservation of symmetry, such as energy-momentum or
global symmetry currents. It is then natural to expect that
hydrodynamics should be emerging in the gravity side of finite
temperature plasma described by black-branes. Indeed, it is based on
this idea that various hydrodynamic coefficients including the
viscosity-entropy ratio
were calculated via linear response theory
in the gravity
background \cite{Policastro:2001yc,Kovtun:2003wp,Buchel:2003tz,Natsuume:2007ty,David:2009np,Hartnoll:2007ip,Baier:2007ix}.
{\it Ab initio} way of deriving the
hydrodynamics from the gravity side when black-brane parameters like
horizon and charges are slowly varying was
recently developed \cite{Bhattacharyya:2008jc,Bhattacharyya:2008ji,Haack:2008cp,Bhattacharyya:2008mz}, and
extended to the single $U(1)$ R-charged system in ref.\cite{Erdmenger:2008rm,Banerjee:2008th,Hur:2008tq}.
This progress
is important because one can in principle go beyond the linearized
approximation to arbitrary non-linear orders one desires, and some
of high order transport coefficients and non-linearity seem
interesting \cite{Haack:2008xx}.
For applications to non-relativistic AdS/CFT, see ref.\cite{Bhattacharyya:2008kq,Rangamani:2008gi,Herzog:2008wg},
and for dyonic system, see ref.\cite{Hansen:2008tq,Caldarelli:2008ze}. See also ref.\cite{Kanitscheider:2009as,Fouxon:2008ik}
for similar developments.

In this work, we generalize this line of development in two
different ways, motivated by real QCD plasma. In QCD with several
quark flavors, there is an enlarged global symmetry $SU(N_f)_L\times
SU(N_f)_R\times U(1)_B$. If one first focuses on the quark species,
each quark flavor has its own conservation; in the case of three
quarks $u,d,s$ that seem relevant in the RHIC experiment, one should
deal with finite temperature plasma with $U(1)^3$ global symmetry.
Note that these $U(1)^3$ components are highly interacting with each
other by strong interactions, and except their conservation laws one
can not predict {\it a priori} anything about their dynamics such as diffusion
coefficient etc. Because these interactions are crucially affecting
the results of transport coefficients, we better work in a
well-defined AdS/CFT set-up rather than working in an arbitrary
unguided gravity theory. We choose to study the STU model, which is
a consistent truncation of $AdS_5\times S^5$ with $U(1)^3$ (or any
Toric Sasaki-Einstein compactification) dual to $N=4$ SYM plasma
with three Cartan $U(1)$'s inside $SO(6)_R$, as a model example of
multi-charged finite temperature plasma. This model has been previously studied in linear response approach
in ref.\cite{Mas:2006dy,Son:2006em}, but our framework enables one to go beyond linearized approximation.
We also compute charge diffusion coefficients in the model for the first time in the literature.
Although our set-up is not
precisely QCD, we hope that it captures some aspect of real $U(1)^3$
dynamics in RHIC plasma.

Our second generalization is for non-Abelian $SU(2)$, although our
result is valid for an arbitrary Lie group\footnote{For study of $SU(2)$ in a different context
of condensed matter system, See ref.\cite{Gubser:2008zu,Herzog:2009ci,Ammon:2009fe}.}. This has a clear
motivation from QCD again; it mimics iso-spin $SU(2)_V$ symmetry of
QCD in two-flavor approximation in a late stage of RHIC plasma. As
fluctuations of charged pion density correspond to $SU(2)_V$
fluctuations, our study should be interesting in describing pion
fluid at finite temperature too. For a specific model, we study a
realization of $SU(2)$ gauged supergravity in Tri-Sasakian
compactification of M-theory to $AdS_4$, corresponding to the
$SU(2)_R$-symmetry sector of 3-dimensional M2-brane plasma. However
as we are more interested in higher dimensions such as 4-dimensions,
we simply generalize the bosonic action to arbitrary dimensions
$AdS_{n+1}$ with $n\ge 3$ and perform analysis in complete generality of
dimensions. Our analysis automatically includes the generalization
of the previous single $U(1)$ case to arbitrary dimensions as well.
One should however note that our $U(1)^3$ and $SU(2)$ R-symmetry in the field theory side
do not come from fundamental flavors, but rather from flavors with adjoint representation. To
introduce fundamental flavors, one normally needs to introduce extra branes in the set-up, whose
detailed study in hydrodynamics is remained for future work.

We obtain the results at first order in derivatives, while the
second order calculation is straightforward, as we will mention in
the text, but extremely complicated to present. We leave its more
controlled analysis to the future. We however mention that even the
first order transport coefficients we obtain are quite non-trivial.
In the STU model, we find three conserved currents of each $U(1)^3$ at first order in derivatives to be
\be
J^\mu_{I}= \rho_I u^\mu  -{\cal D}_I \left(\eta^{\mu\nu}+u^\mu u^\nu\right) D_\nu \rho_I +\zeta_I
\epsilon^{\nu\rho\sigma\mu}u_\nu\partial_\rho u_\sigma+\cdots \quad,
\ee
where $I=1,2,3$ runs for $U(1)^3$ symmetries, $\rho^I$ is the charge density, and the diffusion coefficients ${\cal D}_I$ are given by
\be
{\cal D}_I = {(r_H^2-q_I)\over 2 r_H^3 H^{1\over 2}(r_H)}\quad,
\ee
with $r_H$ being the horizon radius, and the relation of $q_I$ with the energy/charge densities can be easily found in the text.
The parity-violating coefficients $\zeta_I$ originated from the 5D Chern-Simons term are
\bear
\zeta_I &=& {1\over 32 \pi G_5}\Bigg( C_{IJK} {\sqrt{m q_J} \sqrt{m q_K}
\over (r_H^2+q_J)(r_H^2+q_K)} \nonumber\\
&-& {\sqrt{m q_I}\over 3m} C_{JKL} {\sqrt{m q_J}\sqrt{m q_K} \sqrt{m q_L}
\over (r_H^2+q_J)(r_H^2+q_K)(r_H^2+q_L)}\Bigg) \quad.
\eear
where $C^{IJK}$ is the Chern-Simons coefficient.
As far as we know, this is the first time in literature to have these results.
For the $SU(2)$ case, our result for the first order correction to the $SU(2)$ currents in $n$-dimensional CFT is
\bear
J^{a(1)}_\mu &=&
-{\cal D} \left({\rho\cdot P^\nu_\mu (\partial_\nu\rho)\over \rho\cdot\rho}-u^\nu\partial_\nu u_\mu\right)
\rho^a+{\cal D}_1 \epsilon^{abc}\rho^b P^\nu_\mu (\partial_\nu \rho^c)\nonumber\\
&+&{\cal D}_2P^\nu_\mu\left(\rho^a (\rho\cdot
\partial_\nu\rho)-(\rho\cdot\rho) (\partial_\nu\rho^a) \right)\quad,
\eear with three diffusion coefficients\bear {\cal D} &=& {1\over
(n-2) r_H}\left(1-{2 \vec q\cdot \vec q \over n m r_H^{n-2}}\right)={(n-2)m+2 r_H^n \over n(n-2)m r_H} \quad,\nonumber\\
 {\cal D}_1&=& {8\pi G_{n+1} f^{(n-2)}\over
(n-1)} \quad,\quad {\cal D}_2={2^{11\over 2} \pi^2 G_{n+1}^2
g^{(n-2)}\over (n-1)^{3\over 2}(n-2)^{1\over 2}}\quad. \eear The
${\cal D}$ is essentially the usual diffusion coefficient of Abelian
nature, while the other two diffusion coefficients are due to the
non-Abelian properties. The constants $f^{(n-2)}$ and $g^{(n-2)}$ are defined in the text.
We hope that this structure is a useful starting point to study non-Abelian iso-spin plasma of RHIC.

\section{Crash review of the method}

The basic idea in ref.\cite{Bhattacharyya:2008jc,Bhattacharyya:2008ji} for deriving hydrodynamics from gravity is
conceptually quite neat; given a black-brane solution with certain
parameters such as temperature, charges etc, one simply considers
slowly-varying those parameters in the solution. Note that keeping
the form of the original solution while only varying the parameters
inside has a physical meaning of {\it local} thermal equilibrium
with given parameters at that point. However, the resulting
configuration with varying parameters will no longer solve the
equations of motion by obvious reason, and one should add
corrections to the original form of the solution to satisfy the
equations of motion. These corrections will clearly be sourced by
the derivatives of the black-brane parameters one is slowly-varying,
and one can systematically invoke derivative expansion for these
corrections. After obtaining the full solution at $k$-'th order in
derivatives, one can read off physical quantities at that order via
AdS/CFT dictionary, such as energy-momentum tensor and charge
currents. In principle one can go to an arbitrary order in
derivative expansion systematically.

We simply illustrate the procedure for the metric and we refer to
the original work of ref.\cite{Bhattacharyya:2008jc,Bhattacharyya:2008ji} for more complete discussion. Suppose we
have a homogeneous black-brane solution with the metric
$g^{(0)}_{MN}(u_\mu,m,Q_i)$ where $u_\mu$, $m$ and $Q_i$'s are
parameters of the solution such as 4-velocity, energy density and
charge density of the plasma. Once we allow for these parameters to
vary in CFT coordinate directions up to $k$'th derivative, one
should add correction terms
\be
g_{MN}=g^{(0)}_{MN}\left(u_\mu(x),m(x),Q_i(x)\right)+\sum_{i=1}^k g^{(i)}_{MN}\quad,
\ee
to solve the equations of motion, where $u_\mu(x),\dots$ are varying in CFT coordinate $x^\mu$
up to $k$'th derivatives only, and $g^{(i)}_{MN}$ is a local function of derivatives
at $i$'th order. Suppose one solved this problem up to $k$'th order. Then to go to the next order,
one simply considers the varying parameters {\it in the above $k$'th solution} up to $(k+1)$'th order,
which would not solve the equation of motion any more due to $(k+1)$'th order in derivatives we
are now considering. Because the equations of motion are solved up to $k$'th order already by
the above, one simply needs to add $g^{(k+1)}_{MN}$ to $g_{MN}$ that is a local function of
$(k+1)$'th order in derivatives. Typically the equations of motion are second order
partial differential equations, and since the variation of $g^{(k+1)}_{MN}$ itself along $x^\mu$
would be the next order to be neglected, $g^{(k+1)}_{MN}$ only depends on the holographic direction $r$
and the {\it local} derivatives of parameters, without any $x^\mu$ dependence at this order.
Therefore one would get a simple second order ordinary differential equation along $r$ direction from the equations of motion,
\be
L_r \left(g^{(k+1)}_{MN}\right) = S_{MN}^{(k+1)}\quad,
\ee
with a source $S_{MN}^{(k+1)}$ being some function of {\it local} derivatives at $(k+1)$'th order.
Note that $L_r$ would be universal, without being dependent on $k$, completely determined
by the zero'th order solution, and one can obtain the source
$S_{MN}^{(k+1)}$ quite straightforwardly by plugging the above $k$'th order solution {\it with
$(k+1)$'th order derivatives of parameters} into the equation of motion, and simply gathering uncanceled
$(k+1)$'th order term left. Therefore even after performing the analysis at the first order,
one can find the ordinary differential operator $L_r$, and the subsequent higher order analysis
will then become conceptually simpler.
One can go on these steps inductively to arbitrary order in derivatives.

\section{Hydrodynamics with $U(1)^3$ : the STU model}

The action of the STU model which is a sub-sector of $AdS_5$ gauged $N=4$ supergravity holographically
dual to $N=4$ SYM theory is
\bear
(16\pi G_5){\cal L}&=& R\,\,+\,\, 2{\cal V}(X)\,\,-\,\,{1\over 2}G_{IJ}(X)(F^I)_{MN} (F^J)^{MN}
-G_{IJ}(X)\partial_M X^I \partial^M X^J\nonumber\\
&+&{1\over 24\sqrt{-g_5}}\epsilon^{MNPQR} C_{IJK}(F^I)_{MN}(F^J)_{PQ}(A^K)_R\quad,
\eear
where
\be
{\cal V}(X)= 2 \sum_{I=1}^3 {1\over X^I}\quad,\quad G_{IJ}={1\over 2}{\rm diag}\left({1\over (X^I)^2}\right)
\quad,
\ee
and $C_{IJK}$ is totally symmetric with $C_{123}=1$. Also, $X^I$ are not independent but
constrained by
\be
{1\over 6} C_{IJK} X^I X^J X^K = X^1 X^2 X^3 =1\quad.
\ee
We have put $L^4=4\pi g_s N l_s^4 \equiv 1$ for simplicity and in this convention, we have
\be
G_5={\pi\over 2 N^2}\quad,
\ee
where $N$ is the rank of the gauge group.
Capital letters $M,N,\cdots$ represent 5-dimensional indices, while Greek letters
$\mu,\nu,\cdots$ would mean 4-dimensional indices.
The equations of motion one obtains consist of the Einstein equation
\bear
R_{MN}&+&\left({2\over 3}{\cal V}(X)\,\,+{1\over 6}G_{IJ}(X)(F^I)_{MN}(F^J)^{MN} \right)g_{MN}\nonumber\\
&-&G_{IJ}(X)(F^I)_{PM}{(F^J)^P}_N \,\,-\,\,G_{IJ}(X)\partial_M X^I \partial_N X^J \,\,=\,\,0\quad,
\eear
the three Maxwell equations for each $I=1,2,3$,
\be
\nabla_N\left(G_{IJ}(X)(F^J)^{MN}\right)\,\,-\,\,{1\over 16\sqrt{-g_5}}\epsilon^{MNPQR}
C_{IJK}(F^J)_{NP}(F^K)_{QR}\,\,=\,\,0
\ee
and the scalar field equations
\bear
&&\left(\nabla_M\left(G_{IJ}(X)\partial^M X^J \right) +{\partial{\cal V}(X)\over \partial X^I}
\right){\delta X^I \over \delta \phi^i}\nonumber\\
&& -{1\over 2}\left({\partial G_{IJ}(X)\over \partial X^K}\right)\left(\partial_M X^I \partial^M X^J +{1\over 2} (F^I)_{MN}
(F^J)^{MN} \right){\delta X^K\over \delta\phi^i} =0\quad,
\eear
where $\phi^i$ ($i=1,2$) are any independent parametrization of $X^I$'s.

The black brane solution with arbitrary three charges has been known \cite{Behrndt:1998jd} and is given by
\footnote{The gauge fields in the solution have an infinite norm
at the horizon due to the diverging $g^{00}$, which can be remedied by going to the grand-canonical ensemble
with chemical potential that can be added to the solution as a constant mode. We thank Mukund Rangamani for pointing
this issue to us.}
\bear
ds^2 &=& -H^{-{2\over 3}}(r)f(r) u_\mu u_\nu dx^\mu dx^\nu
-2 H^{-{1\over 6}}(r) u_\mu dx^\mu dr + r^2 H^{1\over 3}(r)  \left(\eta_{\mu\nu}+u_\mu u_\nu\right)dx^\mu dx^\nu
\nonumber\\
A^I &=& {\sqrt{m q_I}\over r^2+q_I} u_\mu dx^\mu \quad,\quad
X^I \,\,=\,\, {H^{1\over 3}(r) \over H_I(r)}\quad, \label{0thsol}
\eear
where
\be
f(r)=-{m\over r^2}+r^2 H(r)\quad,\quad H(r)=\prod_{I=1}^3 H_I(r)\quad,\quad H_I(r)=1+{q_I\over r^2}\quad,
\ee
and $u_\mu$ is the 4-velocity of the fluid with $u_\mu u^\mu=-1$. Our convention is
$\eta_{\mu\nu}={\rm diag}(-,+,+,+)$ and $u_\mu=(-1,0,0,0)$ in the rest frame.

As we discussed in the previous section,
we consider slowly varying parameters $u_\mu$, $m$, and $q_I$ up to first order, and
we work in the frame where $u_\mu=(-1,0,\cdots,0)$ at the position $x^\mu=0$ for simplicity.
Once we find the solution, we can easily make the result relativistically covariant.
Then at first order in derivatives, we have
\bear
u_\mu&=&(-1, x^\mu\partial_\mu u_i)\nonumber\\
m&=& m^{(0)}+x^\mu\partial_\mu m\nonumber\\
q_I&=& q_I^{(0)}+x^\mu\partial_\mu q_I\quad,
\eear
and the above black-brane solution will no longer be a solution with these varying parameters.
To be a solution, we have to add the corrections $g^{(1)}_{MN}$, $A^{I(1)}_M$ and $X^{I(1)}$
to the zero'th order solution with varying parameters,
which should be chosen to satisfy the equations of motion. These corrections
will be proportional to the first derivatives of the varying parameters, and we can
neglect their variations along $x^\mu$ as it would be second order, so these corrections
$g^{(1)}_{MN}$, $A^{I(1)}_M$ and $X^{I(1)}$ are functions only on the $r$-coordinate.
One can choose the gauge using coordinate re-parametrization and gauge transformations
to be\footnote{The last gauge is different from the one in ref.\cite{Bhattacharyya:2008jc,Bhattacharyya:2008ji,Bhattacharyya:2008mz,Banerjee:2008th},
but same as in ref.\cite{Haack:2008cp,Erdmenger:2008rm}}
\be
g^{(1)}_{rr}=0\quad,\quad g^{(1)}_{r\mu}\sim u_\mu\quad,\quad A^{I(1)}_r=0\quad,
\quad \sum_{i=1}^{3}g^{(1)}_{ii}=0\quad.
\ee

Let us write the 0'th order metric as
\be
ds^2 = -A(r)dt^2 +2 B(r)dt dr +C(r) (dx^i)^2\quad,
\ee
with
\be
A(r)=H^{-{2\over 3}}(r)f(r)\quad,\quad B(r)= H^{-{1\over 6}}(r)\quad,\quad C(r)=r^2 H^{1\over 3}(r)  \quad.
\ee
Then the metric up to first order in derivatives including the correction $g^{(1)}_{MN}$
looks as
\bear
ds^2 &=& -A(r)dt^2 +2 B(r)dt dr +C(r) (dx^i)^2\nonumber\\
   &+&\left[-x^\mu\left(\partial_\mu A\right) +g^{(1)}_{tt}(r)\right] dt^2
   +2\left[x^\mu\left(\partial_\mu B\right)+g^{(1)}_{tr}(r)\right]dt dr
   +2\left[-x^\mu\left(\partial_\mu u_i\right) B(r)\right]dr dx^i\nonumber\\
   &+&2\left[x^\mu\left(\partial_\mu u_i\right)(A(r)-C(r)) +g^{(1)}_{ti}(r)\right] dt dx^i
   +\left[x^\mu\left(\partial_\mu C\right)\delta_{ij} +g^{(1)}_{ij}(r)\right] dx^i dx^j\quad,
\eear
and the gauge fields become
\bear
A^I&=&-{\sqrt{m q_I}\over r^2+q_I}dt \nonumber\\
&+&\left[-x^\mu\partial_\mu\left(\sqrt{m q_I}\over r^2+q_I\right) +A^{I(1)}_t(r)\right] dt
+\left[x^\mu \left(\partial_\mu u_i\right){\sqrt{m q_I}\over r^2+q_I} +A^{I(1)}_i(r)\right] dx^i,
\eear
and finally scalar fields will be
\be
X^I={H^{1\over 3}(r) \over H_I(r)}+x^\mu \left(\partial_\mu X^I\right) +X^{I(1)}(r)\quad.
\ee
The task is to insert the above into the original equations of motion to obtain the equations for
the first order corrections $g^{(1)}_{MN}$, $A^{I(1)}_M$ and $X^{I(1)}$, and then to solve them.

Because we have a spatial $SO(3)$ symmetry in our rest frame, one finds that the equations for
the first order corrections decompose into different $SO(3)$ representations.
After a tedious but straightforward calculation, one finds the following ordinary differential
equations for $g^{(1)}_{MN}$, $A^{I(1)}_M$ and $X^{I(1)}$.
The easiest one is the tensor mode, that is, traceless $ij$ components of the Einstein equation,
\be
-{1\over 2r}\partial_r\left(r^3f(r)\partial_r\left(g_{ij}^{(1)}\over r^2 H^{1\over 3}(r)\right)\right)
= {1\over 2r} \partial_r\left( r^3 H^{1\over 2}(r) \right) \left(\partial_i u_j +\partial_j u_i -{2\over 3}
\left(\partial_k u_k\right) \delta_{ij}\right)\quad.\label{tensor}
\ee

The vector mode equations are more complicated. From $ti$ component of the Einstein equation, one has
\bear
&&-{f(r)\over 2 r^3 H(r)  }\partial_r\left(  r^5 H(r)  \partial_r\left(g^{(1)}_{ti}\over  r^2 H^{1\over 3}(r)
\right)\right)
-\sum_{I=1}^3 {f(r)\sqrt{m q_I}\over r^3 H(r)}\left(\partial_r A^{I(1)}_i\right)\nonumber\\
&&= {f(r)\over H^{1\over 2}(r)}\left( {2m\over r^3 f(r)} +{1\over 2r}\sum_{I=1}^3 {1\over H_I(r)}\right)
\left(\partial_t u_i\right)+{f(r)\over H^{1\over 2}(r)}\left({\left(\partial_i m\right)
\over 2 r^3 f(r) } - P^{(1)}_i\right)\quad,\label{vectorti}
\eear
and from the $ri$-component,
\bear
&&{1\over 2 r^3 H^{1\over 2}(r)}\partial_r\left(r^5 H(r) \partial_r\left(
g^{(1)}_{ti}\over  r^2 H^{1\over 3}(r)\right)\right) +\sum_{I=1}^3
{\sqrt{m q_I}\over r^3 H^{1\over 2}(r)}\left(\partial_r A^{I(1)}_i\right)\nonumber\\
&&=-{1\over 2r}\left(\sum_{I=1}^3 {1\over H_I(r)}\right)\left(\partial_t u_i\right) +P^{(1)}_i\quad,\label{vectorri}
\eear
where $P^{(1)}_i$ is a complex expression in terms of first order spatial derivatives in $q_I$'s only,
which is given in the Appendix.
From the $i$-component of the Maxwell equation for each $I$, we have
\bear
&&-{1\over r}\partial_r\left({r f(r)\left(H_I(r)\right)^2\over H(r)}\left(\partial_r A^{I(1)}_i\right)\right)
-{2\sqrt{m q_I}\over r}\partial_r\left(g^{(1)}_{ti}\over  r^2 H^{1\over 3}(r)
\right)\nonumber\\
&& ={1\over r}\partial_r\left({\sqrt{m q_I} H_I(r)\over r H^{1\over 2}(r)}\left(\partial_t u_i\right)\right)
+{1\over r}\partial_r
\left(-{1\over 2}C_{IJK}{\sqrt{mq_J}\sqrt{m q_K}\over (r^2+q_J) (r^2+ q_K)}\epsilon^{ijk}\left(
\partial_j u_k\right)\right)\nonumber\\
&&
+{1\over r}\partial_r\left({1\over 2 r^3 H^{1\over 2}(r)\sqrt{m q_I}}\left(m(r^2-q_I)
\left(\partial_i q_I\right) + q_I(r^2+q_I)\left(\partial_i m\right)\right)\right)\quad.\label{vectori}
\eear
The above five equations are vector mode equations. We mention that the ordinary differential operators
in the left-hand side of the equations in the above and below take  {\it integrable} forms, which is a quite non-trivial
fact that has been checked via complicated algebra and
educated guesses\footnote{In verifying these, we sometimes used Mathematica for basic algebra manipulations.}.

Finally, the most complicated part is the scalar mode equations under $SO(3)$.
From the $tt$-component of Einstein equation, one has
\bear
&&-{f(r)\over 2r^3 H(r)}\partial_r\left(r^3 H^{2\over 3}(r)\partial_r g^{(1)}_{tt}\right)
-{4\over 3}\sum_{I=1}^3 {f(r)\sqrt{m q_I}\over r^3 H(r)}\left(\partial_r A^{I(1)}_t\right)
\nonumber\\
&&-{f(r)\over 2H^{1\over 3}(r)}\partial_r\left(H^{-{2\over 3}}(r)f(r)\right)\partial_r\left(H^{1\over 6}(r)
g^{(1)}_{tr}\right)-{8\over 3}{f(r)\sum_{I=1}^3 H_I(r)\over  H(r)}
\left(H^{1\over 6}(r)
g^{(1)}_{tr}\right)\nonumber\\
&&+{4\over 3}{f(r)\over r^4 H^{4\over 3}(r)}\sum_{I=1}^3 \left((r^2+q_I)^2+{2 m q_I\over r^2+q_I}\right)X^{I(1)}
=S^{(1)}_{tt}\quad,
\eear
from the $tr$-component,
\bear
&&{1\over 2 r^3 H^{1\over 2}(r)}\partial_r\left(r^3 H^{2\over 3}(r)\partial_r g^{(1)}_{tt}\right)
+{4\over 3}\sum_{I=1}^3 {\sqrt{mq_I}\over r^3 H^{1\over 2}(r)}\left(\partial_r A^{I(1)}_t\right)\nonumber\\
&&+{1\over 2} H^{1\over 6}(r)\partial_r\left(H^{-{2\over 3}}(r)f(r)\right)
\partial_r\left(H^{1\over 6}(r)
g^{(1)}_{tr}\right)+{8\over 3}{\sum_{I=1}^3 H_I(r)\over  H^{1\over 2}(r)}
\left(H^{1\over 6}(r)
g^{(1)}_{tr}\right)\nonumber\\
&&-{4\over 3}{f(r)\over r^4 H^{5\over 6}(r)}\sum_{I=1}^3 \left((r^2+q_I)^2+{2 m q_I\over r^2+q_I}\right)X^{I(1)}
=S^{(1)}_{tr}\quad,
\eear
the $rr$-component looks as
\bear
\partial_r\left(\log\left(r^3 H^{1\over 2}(r)\right)\right) \partial_r\left(
H^{1\over 6}(r) g^{(1)}_{tr}\right)
-\sum_{I=1}^3 \partial_r\left(\log\left(H^{1\over 3}(r)\over H_I(r)\right)\right)
\partial_r\left({H_I(r)\over H^{1\over 3}(r)} X^{I(1)}\right)=0,
\eear
and the last scalar mode
equation from the Einstein equation is the trace part, that is $\sum_{i=1}^3 (ii)$,
\bear
&&{3\over 2r}\partial_r\left(r H^{1\over 3}(r)\partial_r\left(r^2 H^{1\over 3}(r)\right) g^{(1)}_{tt}
\right)
-{2\over r}\sum_{I=1}^3 \sqrt{mq_I}\left(\partial_r A^{I(1)}_t\right)\nonumber\\
&&+{3\over 2}{\partial_r\left(r^2H^{1\over 3}(r)\right) f(r)\over H^{1\over 3}(r)}
\partial_r\left(H^{1\over 6}(r) g^{(1)}_{tr}\right)
+8r^2\left(\sum_{I=1}^3 H_I(r)\right)\left(H^{1\over 6}(r) g^{(1)}_{tr}\right)\nonumber\\
&&-{4\over r^2 H^{1\over 3}(r)}\sum_{I=1}^3 \left((r^2+q_I)^2-{m q_I\over r^2+q_I}\right) X^{I(1)}=
\sum_{i=1}^3S_{ii}^{(1)}\quad,
\eear
where $S^{(1)}_{tt}$, $S^{(1)}_{tr}$, and $\sum_{i=1}^3 S^{(1)}_{ii}$ are source terms
proportional to space-time derivatives of black-brane parameters, which are given in the Appendix.

One obtains more scalar mode equations from the Maxwell equations. From the $t$-part of the Maxwell
equations for each $I$,
one gets
\bear
&&{2 f(r)\sqrt{mq_I}\over r^3 H(r)}\partial_r\left(H^{1\over 6}(r) g^{(1)}_{tr}\right)
-{f(r)\over r^3 H(r)}\partial_r\left(r^3 H^2_I(r)\partial_r A^{I(1)}_t\right)\nonumber\\
&&+{4 f(r)\sqrt{mq_I}\over r^3 H(r)}\partial_r\left({H_I(r)\over  H^{1\over 3}(r)} X^{I(1)}\right)
={2\over r^3 H^{1\over 2}(r)}\bigg(\partial_t\left(\sqrt{mq_I}\right)+\sqrt{mq_I}\left(\partial_i
u_i\right)\bigg)\quad,
\eear
and from the $r$-component,
\bear
&& {-2\sqrt{mq_I}\over r^3 H^{1\over 2}(r)}\partial_r\left(H^{1\over 6}(r) g^{(1)}_{tr}\right)
+{1\over r^3 H^{1\over 2}(r)}\partial_r\left(r^3 H^2_I(r)\partial_r A^{I(1)}_t\right)\nonumber\\
&&-{4\sqrt{mq_I}\over r^3 H^{1\over 2}(r)}\partial_r\left({H_I(r)\over H^{1\over 3}(r)}X^{I(1)}\right)=0\quad.
\eear
Lastly, we present the scalar field equations. For this purpose, we choose $X^1$ and $X^2$ as independent
variables with $X^3={1\over X^1 X^2}$. We have
\bear
&&-{1\over r^3 H^{1\over 3}(r)}\partial_r\left(r^3 H^{{2\over 3}}(r)\partial_r\left(
\log\left(H^{1\over 2}(r)\over H_1(r)H^{1\over 2}_2(r)\right)\right) g^{(1)}_{tt}\right)\nonumber\\
&&
-{2\over r^3 H^{1\over 3}(r)}\left(\sqrt{mq_1}\left(\partial_r A^{1(1)}_t\right) -\sqrt{mq_3}
\left(\partial_r A^{3(1)}_t \right)\right)\nonumber\\
&&
-{f(r)\over H^{1\over 3}(r)}
\left(
\log\left(H^{1\over 2}(r)\over H_1(r)H^{1\over 2}_2(r)\right)\right)\partial_r\left(
H^{1\over 6}(r) g^{(1)}_{tr}\right)-{4(q_1-q_3)\over r^2 H^{1\over 3}(r)}\left(H^{1\over 6}(r) g^{(1)}_{tr}\right)
\nonumber\\
&& +{1\over r^3 H^{1\over 3}(r)}\partial_r\left(r^3 f(r)\partial_r\left({H_1(r)\over H^{1\over 3}(r)} X^{1(1)}
+{H_2(r)\over 2 H^{1\over 3}(r)}X^{2(1)}\right)\right)\nonumber\\
&&+{2\over r^2 H^{1\over 3}(r)}\left((r^2+q_1)+{2 mq_1\over (r^2+q_1)^2}\right)\left({H_1(r)\over H^{1\over 3}(r)}
X^{1(1)}\right)\nonumber\\
&&+{2\over r^2 H^{1\over 3}(r)}\left((r^2+q_3)+{2 m q_3\over (r^2+q_3)^2}\right)\left(
{H_1(r)\over H^{1\over 3}(r)}X^{1(1)}+{H_2(r)\over H^{1\over 3}(r)}X^{2(1)}\right)=S^{1(1)},
\eear
and the similar equation with $X^1$ and $X^2$ interchanged. The source terms $S^{1(1)}$ and $S^{2(1)}$
can again be found in the Appendix.
In the Appendix, we also sketch our method of computations for deriving these equations,
the main task being to obtain the variation of Ricci tensor up to first order.

The main point in this heavy endeavor
is in fact to solve the above equations to find the first order
transport coefficients; luckily enough, we are able to solve the above equations {\it in explicit integral forms}.

\subsection{The solution}

We first observe that some combinations of the above equations are in fact {\it constraints}
on the space-time derivatives of the black-brane parameters; they can't be arbitrary but have
to be consistent with the conservations laws in the CFT side, such as energy-momentum and current
conservations. In other words, bulk equations of motion {\it include} the conservation laws
in the CFT side. Writing the Einstein equation as $E_{MN}$ and the Maxwell equations as $M^I_M$,
one has three kinds of constraints equations,
\bear
0&=&g^{rt} E_{tt}+g^{rr}E_{rt}=-\left(H^{1\over 6}(r) S^{(1)}_{tt}+H^{-{1\over 3}}(r)f(r)S^{(1)}_{tr}\right)\nonumber\\
&=&{-1\over 2 r^9 H^{3\over 2}(r)}\Bigg(
\left(2mq_1q_2q_3  \left(\partial_i u_i\right)+m\partial_t\left(q_1q_2q_3\right)\right)\nonumber\\
&-&\left(\left(\partial_t m\right)\left(q_1q_2+q_2q_3+q_3q_1\right)-m\partial_t\left(
q_1q_2+q_2q_3+q_3q_1\right)\right)r^2\nonumber\\
&-& \left(2 m\left(\partial_i u_i\right)\left(q_1+q_2+q_3\right)+2\left(\partial_t m\right)
\left(q_1+q_2+q_3\right)-m\partial_t\left(q_1+q_2+q_3\right)\right)r^4 \nonumber\\ &-&
\left(4m\left(\partial_i u_i\right)+3\left(\partial_t m\right)\right)r^6\Bigg)\quad,\\
0&=& g^{rt}M_t^I +g^{rr} M_r^I = {-2\over r^3 H^{1\over 3}(r)}
\bigg(\partial_t\left(\sqrt{mq_I}\right)+\sqrt{mq_I}\left(\partial_i
u_i\right)\bigg)\quad,\\
0&=& g^{rt} E_{ti}+g^{rr} E_{ri}={-1\over r^3 H^{1\over 3}(r)} \left(
{1\over 2}\left(\partial_i m\right)+2m\left(\partial_t u_i\right)\right)\quad.
\eear
Note that for the above combinations of equations of motions, the radial differential operators
cancel with each other to leave the above algebraic constraints.
The constraints are uniquely solved by
\bear
\left(\partial_t m\right) &=& -{4\over 3}m \left(\partial_i u_i\right)\quad,
\quad \left(\partial_i m\right)= -4 m \left(\partial_t u_i\right)\quad,\nonumber\\
\left(\partial_t q_I\right)&=& -{2\over 3} q_I\left(\partial_i u_i\right)\quad {\rm or\,\,equivalently,}\quad
\partial_t \left(\sqrt{mq_I}\right) = -\sqrt{mq_I}\left(\partial_i u_i\right)\quad,\label{constr}
\eear
where the first two equations imply the {\it zero'th order} energy momentum conservation, and
the last equation is the conservation of $U(1)^3$ global symmetry currents.
Indeed, applying AdS/CFT dictionary to our zero'th order black-brane solution, we have
\bear
T^{\mu\nu(0)}&=&     {m\over 16\pi G_5}\left(\eta^{\mu\nu} +4 u^\mu u^\nu\right)\equiv
p\left(\eta^{\mu\nu} +4 u^\mu u^\nu\right)
\quad,       \nonumber\\
J_I^{\mu(0)} &=& {\sqrt{mq_I}\over 8\pi G_5}u^\mu  \equiv  \rho_I u^\mu\quad,\label{0thcurrent}
\eear
whose conservation laws in our rest-frame $u_\mu=(-1,0,0,0)$ are nothing but (\ref{constr}).
One generically obtains conservation laws of $(k-1)$'th order from the $k$'th order equations of motion.

We next solve the remaining dynamical equations.
It is easiest to solve the tensor mode equation (\ref{tensor}).
Integrating it gives us
\be
g^{(1)}_{ij}(r)=r^2 H^{1\over 3}(r)\left(
-2\sigma_{ij}\int^r_{\infty} dr'\,{H^{1\over 2}(r')\over f(r')}+C_{ij}\int^r_\infty dr'
\,{1\over r'^3 f(r')}+C_{ij}'\right)\quad,
\ee
with some constants $C_{ij}$ and $C_{ij}'$, and we define
\be
\sigma_{ij} ={1\over 2}\left(\partial_i u_j +\partial_j u_i -{2\over 3}
\left(\partial_k u_k\right) \delta_{ij}\right)\quad.
\ee
One has to put $C_{ij}'=0$ as it is a non-normalizable mode. The $C_{ij}$ is uniquely determined
to give a regular solution at the horizon $r=r_H$ where $f(r_H)=0$; note that the two integrals
in the above logarithmically diverge near $r=r_H$, and having a cancelation between the two
for a finite result uniquely
fixes
\be
C_{ij}= 2r_H^3 H^{1\over 2}(r_H)\sigma_{ij}\quad.
\ee

The vector mode equations (\ref{vectorti}), (\ref{vectorri}), and (\ref{vectori}) are harder.
As we already solved one constraint equation from (\ref{vectorti}) and (\ref{vectorri}),
we need to solve only (\ref{vectorri}) and (\ref{vectori}).
Let us first integrate (\ref{vectori}) once, which gives us
\bear
&&{r f(r) \left(H_I(r)\right)^2\over H(r)}\left(\partial_r A_i^{I(1)}\right)
+2\sqrt{mq_I} \left(g^{(1)}_{ti}\over r^2 H^{1\over 3}(r)\right)\nonumber\\
&=& -{\sqrt{mq_I} H_I(r)\over r H^{1\over 2}(r)}\left(\partial_t u_i\right)
+{1\over 2}C_{IJK}{\sqrt{mq_J}\sqrt{mq_K}\over (r^2+q_J)(r^2+q_K)} \epsilon^{ijk}\left(\partial_j u_k\right)\label{vector1}
\\
&-&{1\over 2 r^3 H^{1\over 2}(r) \sqrt{mq_I}}\left(m(r^2-q_I)\left(\partial_i q_I\right)
+q_I(r^2+q_I)\left(\partial_i m\right) \right)+ C^I_i \equiv Q^{I(1)}_i(r)+C^I_i\quad,\nonumber
\eear
with some integration constants $C^I_i$, and then
consider the horizon $r=r_H$ where $f(r_H)=0$.
Imposing a regularity on $A_i^{I(1)}$ and $g_{ti}^{(1)}$ at $r=r_H$, the first term drops and we have
\be
\left(g^{(1)}_{ti}(r_H)\over r_H^2 H^{1\over 3}(r_H)\right)
={1\over 2\sqrt{mq_I} } \left(Q^{I(1)}_i(r_H)+C^I_i\right)\quad.
\ee
The point is that the left-hand side is independent of $I$, so that $C^I_i$ can not be arbitrary
but has to take a form
\be
C^I_i= -Q^{I(1)}_i(r_H) +{2\sqrt{mq_I}\over r_H^2 H^{1\over 3}(r_H)} C_i\quad,
\ee
with only one degree of freedom $C_i$, which is nothing but the value of $g^{(1)}_{ti}(r_H)$ at the horizon.
The next step is to use the above (\ref{vector1}) to replace $\left(\partial_r A_i^{I(1)}\right)$
in the equation (\ref{vectorri}) to get a second order differential equation for $g^{(1)}_{ti}$ only;
\bear
&&\partial_r\left(r^5 H(r)\partial_r \left( g^{(1)}_{ti}\over r^2 H^{1\over 3}(r)\right)\right)
-\left(\sum_{I=1}^3 {4mq_I H(r)\over r f(r)\left(H_I(r)\right)^2}\right)
\left( g^{(1)}_{ti}\over r^2 H^{1\over 3}(r)\right)\label{gtieq}\\
&=& \sum_{I=1}^3 {-2\sqrt{mq_I}H(r)\over r f(r) \left(H_I(r)\right)^2}\left(
Q^{I(1)}_i(r)-Q^{I(1)}_i(r_H) +{2\sqrt{mq_I}\over r_H^2 H^{1\over 3}(r_H)} C_i\right)\nonumber\\
&-&r^2 H^{1\over 2}(r)\left(\sum_{I=1}^3 {1\over H_I(r)}\right) \left(\partial_t u_i\right)
+2r^3 H^{1\over 2}(r) P^{(1)}_i(r)\,.\nonumber
\eear

To our surprise, the second order differential operator in the left-hand side is in fact {\it integrable},
that is, the left-hand side can be transformed into
\be
{r^2 H(r)\over f(r)}\partial_r\left({r \left(f(r)\right)^2\over H(r)}\partial_r
\left({H^{2\over 3}(r)\over f(r)} g^{(1)}_{ti}\right)\right)\quad.
\ee
The way we have found this is the following. We start from the Ansatz for an integrable form,
\be
{1\over P}\partial_r\left(Q\partial_r \left(S\cdot \right)\right)=
{1\over P}\left( QS\partial_r^2\cdot +\left(\partial_r\left(QS\right)+Q\left(\partial_r S\right)\right)
\partial_r\cdot + \partial_r\left(Q\partial_r S\right) \cdot \right)\quad,
\ee
and comparing with the left-hand side of (\ref{gtieq}), one gets
\bear
&&{1\over P}QS= r^5 H(r)\quad,\quad {1\over P}\left(\partial_r\left(QS\right)+Q\left(\partial_r S\right)\right)
=\partial_r\left(r^5 H(r)\right)\,,\\
&&{1\over P}\partial_r\left(Q\partial_r S\right)= -\sum_{I=1}^3 {4mq_I H(r)\over r f(r) \left(H_I(r)\right)^2}
\quad.
\eear
One can easily remove $Q$ and $S$ in terms of $P$ to get a differential equation for $P$, which
turns out to be the same differential equation (\ref{gtieq}),
but now without the source term in the right-hand side; that is, $P$ is a homogeneous solution
of the differential operator in (\ref{gtieq}).
Luckily, we know one way to {\it generate} a homogeneous solution without source terms;
recall that any coordinate re-parametrization must correspond to a homogeneous solution of the problem.
For our purpose, the following infinitesimal coordinate transformation
\be
dt\to dt -\epsilon dx^i \quad,\quad dx^i\to dx^i +\epsilon{1\over r^2 H^{1\over 2}(r)} dr\quad,
\ee
generates one homogeneous solution for $g^{(1)}_{ti}$, which is $H^{-{2\over 3}}(r) f(r)$,
and from this one can let $P$ be
\be
P={H^{-{2\over 3}}(r) f(r)\over r^2 H^{1\over 3}(r)} = {f(r)\over r^2 H(r)}\quad.
\ee
Once $P$ is found, it is straightforward to find $Q$ and $S$ from the above equation to have
the above integrable form of the differential operator.

Solving (\ref{gtieq}) then by integrating it once, we have
\bear
&&{r \left(f(r)\right)^2\over H(r)}\partial_r\left({H^{2\over 3}(r)\over f(r) }g^{(1)}_{ti}\right)\label{int1}\\
&&= \int_\infty^r dr'\,\Bigg(
\sum_{I=1}^3 {-2\sqrt{mq_I}\over (r')^3  \left(H_I(r')\right)^2}\left(
Q^{I(1)}_i(r')-Q^{I(1)}_i(r_H) +{2\sqrt{mq_I}\over r_H^2 H^{1\over 3}(r_H)} C_i\right)\nonumber\\
&-&{f(r')\over H^{1\over 2}(r')}\left(\sum_{I=1}^3 {1\over H_I(r')}\right) \left(\partial_t u_i\right)
+{2r'f(r')\over H^{1\over 2}(r')} P^{(1)}_i(r')\Bigg) +C_i'\nonumber\\
&\equiv& \int_{\infty }^r dr'\,I(r')+C_i'\,,\nonumber
\eear
with an integration constant $C'_i$, which can be fixed by considering the behavior at the horizon.
Note that $f(r)$ has an expansion near $r=r_H$ as
\be
f(r)=f'(r_H)(r-r_H)+{1\over 2}f''(r_H)(r-r_H)^2 + \cdots\quad,
\ee
with $f'(r_H)>0$, and the left-hand side in the above (\ref{int1}) takes a limit as $r\to r_H$,
\be
{r \left(f(r)\right)^2\over H(r)}\partial_r\left({H^{2\over 3}(r)\over f(r) }g^{(1)}_{ti}\right)
\to -{r_H f'(r_H)\over H^{1\over 3}(r_H)} g^{(1)}_{ti}(r_H) =
-{r_H f'(r_H)\over H^{1\over 3}(r_H)}C_i\quad,
\ee
where we have used the fact that $g^{(1)}_{ti}(r_H)=C_i$ previously.
This fixes $C'_i$ to be
\bear
C'_i = -\int_\infty^{r_H} dr' I(r')-{r_H f'(r_H)\over H^{1\over 3}(r_H)}C_i\quad,
\eear
where $I(r')$ is the same integrand in (\ref{int1}), and one can rewrite (\ref{int1}) as
\be
{r \left(f(r)\right)^2\over H(r)}\partial_r\left({H^{2\over 3}(r)\over f(r) }g^{(1)}_{ti}\right)
= \int_{r_H}^r dr'\,I(r')-{r_H f'(r_H)\over H^{1\over 3}(r_H)}C_i\quad.
\ee
We then integrate the above once more to have
\bear
g^{(1)}_{ti}(r)&=&
{f(r)\over H^{2\over 3}(r)}\int^r_\infty dr'\,
{H(r')\over r'\left(f(r')\right)^2}\left(\int_{r_H}^{r'} dr''\, I(r'') -{r_H f'(r_H)\over H^{1\over 3}(r_H)}C_i
\right) +{f(r)\over H^{2\over 3}(r)}C_i''\,,\nonumber
\eear
where one needs to put the integration constant $C_i''$ to be zero as it corresponds precisely to
the coordinate re-parametrization that we have used to get a homogeneous solution.

At this point, the only uncertainty we have to fix
is the constant $C_i$ that appears both in $I(r)$ and the above
result. It might seem that it can be fixed for the $g_{ti}^{(1)}$ to have a regular {\it derivative} at the
horizon $r=r_H$; observe that the above result for $g_{ti}^{(1)}$ already has a finite {\it value} at the
horizon irrespective of the constant $C_i$.
Suppose that the integrand in the above equation has an expansion near $r=r_H$,
\be
{H(r')\over r'\left(f(r')\right)^2}\left(\int_{r_H}^{r'} dr''\, I(r'') -{r_H f'(r_H)\over H^{1\over 3}(r_H)}C_i
\right)\sim {a\over(r'-r_H)^2}+{b\over (r'-r_H)}+\cdots\quad,
\ee
then the first piece is harmless after integration as it cancels with $f(r)$ in front, while
the second piece will result in
\be
g^{(1)}_{ti} \sim (r-r_H) \log(r-r_H)+\cdots \quad,
\ee
which has a divergent {\it radial derivative} that would signal  divergent curvature tensors.
Explicitly, one finds that the Ricci tensor $R_{ri}$ diverges with this.
Therefore, we may have to choose right $C_i$ to make sure that $b=0$ in the near horizon expansion.
Explicitly, one has
\be
b= {H(r_H)\over r_H \left(f'(r_H)\right)^2}\left(
I(r_H)- {r_H f'(r_H)\over H^{1\over 3}(r_H)}\left({H'(r_H)\over H(r_H)} -{1\over r_H}
-{f''(r_H)\over f'(r_H)}\right)C_i\right)\quad,
\ee
and moreover one finds from (\ref{int1}) using $f(r_H)=0$ that
\be
I(r_H)=\sum_{I=1}^3{-4 mq_I\over r_H^5 H^{1\over 3}(r_H) \left(H_I(r_H)\right)^2} C_i\quad,
\ee
so that $b$ is in fact proportional to $C_i=g^{(1)}_{ti}(r_H)$, and it appears that we may have to put it zero for regularity. 
However, an explicit computation shows that the coefficient in front of $C_i$ in fact vanishes identically,
and the geometry is smooth for {\it any} value of $C_i$. We need an extra input to fix the constant $C_i$.

The answer to this puzzle lies in the {\it frame choice}, more concretely, the choice of either Landau frame
or Eckart frame. In our work, we choose the Landau frame which states that 
\be
u_\mu T^{\mu\nu} = -\epsilon u^\nu\quad,
\ee 
and in particular, $T^{ti}=0$ must hold in our local rest frame. In holographic renormalization that we will discuss
in more detail in the next section, this condition gives us the constraint that the coefficient
of $1\over r^2$ in near boundary expansion of $g_{ti}^{(1)}(r)$ should vanish, because it is precisely proportional to the first order correction to $T^{ti}$. One can easily check that this indeed fixes our integration constant $C_i$ uniquely.
We have then completely solved for $g^{(1)}_{ti}(r)$, as given above with $C_i$ determined.
Our explicit calculations give us
\bear
C_i&=& {r_H^2 H^{1\over 3}(r_H)\over 4m}\Bigg( \sum_{I=1}^3 \left({4m\over \sqrt{m q_I}}\left({\cal D}_I -{(r_H^2-q_I)\over 2 r_H^2 H^{1\over 2}(r_H)}\right)\left(\partial_i \sqrt{m q_I}\right)\right)\nonumber\\
&+&{\sqrt{m}\over r_H}\left(4r_H^2+3\sum_{J=1}^3 q_J\right)\left(\partial_t u_i\right)\nonumber\\
&+&{1\over 3} C_{IJK}{\sqrt{m q_I}\sqrt{m q_J}\sqrt{m q_K}\over (r_H^2+q_I)(r_H^2+q_J)(r_H^2+q_K)}\epsilon^{ijk}\left(\partial_j u_k\right)\Bigg)\quad,
\eear
where ${\cal D}_I$ is given in the next section.

Once we have the solution for $g^{(1)}_{ti}(r)$, one simply plugs it into
the equation (\ref{vector1}) to solve for $A^{I(1)}_{i}$, whose integration gives us
\be
A^{I(1)}_i=\int_\infty^r dr'\,
{H(r')\over r' f(r') \left(H_I(r')\right)^2}\left(
Q^{I(1)}_i(r')-Q^{I(1)}_i(r_H) -2\sqrt{mq_I}\left({g^{(1)}_{ti}(r') \over (r')^2 H^{1\over 3}(r')}-{C_i\over r_H^2 H^{1\over 3}(r_H)}\right)\right),\label{Asol}
\ee
where we have chosen the integration constant to remove non-normalizable modes.
This completes our long solution for the vector modes under $SO(3)$.

Finally we come to the solution of the scalar modes under $SO(3)$. Although it seems difficult
to our eyes to systematically solve these equations, we are lucky to be able to solve them
by an educated guess from the previous analysis in ref.\cite{Bhattacharyya:2008jc,Banerjee:2008th}; in fact, for the case of single $U(1)$
R-charged hydrodynamics, most of the
scalar modes under $SO(3)$ turn out to be zero, {\it except} $g^{(1)}_{tt}$. Assuming the same
feature, one easily finds that
\be
g^{(1)}_{tt} = {2\over 3} r H^{-{1\over 6}}(r) \left(\partial_i u_i\right)\quad,
\ee
indeed solves all the scalar mode equations of motion
in the previous section\footnote{We used Mathematica for basic algebra manipulations in showing this.}.
As the solution
is expected to be unique up to trivial coordinate re-parametrizations, we
conclude that the above $g^{(1)}_{tt}$ with
\be
g^{(1)}_{tr}=A_t^{I(1)}=X^{I(1)} =0\quad,
\ee
is the solution of the scalar modes under $SO(3)$.

\subsection{The first order transport coefficients}

It is a standard AdS/CFT procedure to obtain the first order corrections to the CFT energy-momentum tensor and
the $U(1)^3$ symmetry currents from the results in the previous section. We first discuss the energy-momentum
tensor. One way to compute the CFT energy-momentum tensor is to rewrite the full first-order metric in the
Fefferman-Graham coordinate,\footnote{It is also possible to get the energy-momentum tensor
directly in the Eddington-Finkelstein coordinate, but the end results should be same
in the first order in derivatives.}
where \be ds^2 = {d\rho^2\over \rho^2} + \rho^2 g_{\mu\nu}(\rho,x)dx^\mu
dx^\nu\quad, \ee and to read off the coefficient of the large $\rho$-expansion of $g_{\mu\nu}(\rho,x)$, \be
g_{\mu\nu}(\rho,x) \sim \eta_{\mu\nu} + \cdots +{ g_{\mu\nu}^{(4)}(x)\over \rho^4}+\cdots\quad. \ee The
holographic renormalization procedure \cite{Bianchi:2001kw}
would then give us \be T_{\mu\nu} = {1\over 4\pi
G_5}g_{\mu\nu}^{(4)}(x)\quad. \label{usualterm}\ee However, if one naively applies this to the zero'th order
black-brane solution (\ref{0thsol}), one {\it does not} find the previously quoted energy-momentum tensor in
(\ref{0thcurrent}), \be T^{\mu\nu(0)}=     {m\over 16\pi G_5}\left(\eta^{\mu\nu} +4 u^\mu u^\nu\right)\quad. \ee
This is due to a subtlety in the scalar fields sector of the STU model; the scalar fields sector provides the
cosmological constant at its vacuum, and hence the boundary counter-term that one adds in the holographic
renormalization involves a non-trivial potential term of the scalar fields $X^I$ \cite{Batrachenko:2004fd}.
Because $X^I$ in the
solution has a non-trivial profile, it turns out that this counter-term gives an additional contribution to the
energy-momentum tensor other than (\ref{usualterm}). The careful analysis including this subtlety was carried
out in ref.\cite{Mas:2006dy} to find the above correct answer.

For our purpose to find the first order correction $T^{\mu\nu(1)}$,
we are however in a lucky situation. Because the first order
corrections to the scalar fields $X^{I(1)}$ vanish, there wouldn't
be any first order contributions from the scalar fields sector to
the energy-momentum tensor, and we can safely use (\ref{usualterm})
for the first order corrections to the energy-momentum tensor. Be
warned that this may not be true in higher orders. A direct
expansion of our first-order solutions in the Fefferman-Graham
coordinate $\rho$ which are related to our $r$ variable by \be
r=\rho+{\left(\partial_i u_i\right)\over 3}-{\left(\sum_I q_I\right)
\over 6 \rho} +{\left(\partial_i u_i\right)\left(\sum_I
q_I\right)\over 54 \rho^2} +\cdots\quad, \ee one finds that the only
non-vanishing first-order correction to the energy-momentum comes
from $g^{(1)}_{ij}$, and is given by \be T^{(1)}_{ij} = -2{r_H^3
H^{1\over 2}(r_H)\over 16\pi G_5} \sigma_{ij}= -2{s\over
4\pi}\sigma_{ij} \equiv -2\eta \sigma_{ij}\quad, \ee where we have
used the fact that the horizon area per unit CFT volume is given by
$r_H^3 H^{1\over 2}(r_H)$, and the entropy density from the
Bekenstein-Hawking formula gives \be s={r_H^3 H^{1\over 2}(r_H)\over
4 G_5}\quad. \ee The last equality is simply the definition of
shear viscosity $\eta$, and one recovers the famous ratio \be
{\eta\over s} ={1\over 4\pi}\quad, \ee in the STU model.
It is not hard to make the previous result in a manifestly covariant standard form away from our
static frame $u_\mu=(-1,0,0,0)$;
\be
T^{\mu\nu} = p\left(\eta^{\mu\nu} +4 u^\mu u^\nu\right)-2\eta\sigma^{\mu\nu}+\cdots \quad,
\ee
with
\be
\sigma^{\mu\nu} = {1\over 2}P^{\mu\alpha}P^{\nu\beta}\left(\partial_\alpha u_\beta+\partial_\beta u_\alpha\right)
-{1\over 3} P^{\mu\nu}\left(\partial_\alpha u^\alpha\right)\quad,
\ee
where $P^{\mu\nu}\equiv \eta^{\mu\nu}+u^\mu u^\nu $ is the projection to the transverse components to $u^\mu$.

It is also possible to write the first-order corrected metric in the covariant form,
\bear
ds^2 &=& - H^{2/3}(r)f(r)u_\mu u_\nu dx^\mu dx^\nu -
2H^{1/6}u_\mu dx^\mu dr+
r^2 H^{1/3}(r)P_{\mu\nu}dx^\mu dx^\nu \cr &+&
\frac{2}{3}rH^{-1/6}(r)(\partial_\rho u_\rho)u_\mu u_\nu dx^\mu
dx^\nu\nonumber\\& -& 2\frac{f(r)}{H^{2/3}(r)}u_\mu\Bigg(
\int_\infty^rdr'\frac{H(r')}{r'f^2(r')}\left(\int_{r_H}^{r'}dr''
I_\nu^{(1)}(r'')-{r_H f'(r_H)\over H^{1/3}(r_H)} C_\mu \right)\Bigg)dx^\mu dx^\nu \cr &+&
2r_H^3H^{1/2}(r_H)r^2 H^{1/3}(r)\sigma_{\mu\nu}\bigg(\int_\infty^r
dr'\frac{1}{f(r')}\bigg(\frac{1}{r'^{3}} -
\frac{1}{r_H^3}\frac{H^{1/2}(r')}{H^{1/2}(r_H)}\bigg)\bigg) dx^\mu
dx^\nu\quad.\qquad
\eear
Here, the covariant first-order radial function
$I_\mu^{(1)}(r)$ is defined as
\bear I^{(1)}_\mu(r) &=&-2\sum_{I=1}^3\frac{\sqrt{mq_I}}{(r)^3H_I^2(r)}\Big(Q_\mu^{I(1)}(r)-Q_\mu^{I(1)}(r_H)\Big)
-\frac{f(r)}{H^{1/2}(r)}u^\nu\partial_\nu
u_\mu\sum_{I=1}^3\frac{1}{H_I(r)} \cr &+&
2\frac{rf(r)}{H^{1/2}(r)}P_\mu^{(1)}(r)\quad,
\eear
where the first-order
functions $Q^{I(1)}_\mu(r)$ and $P^{I(1)}_\mu(r)$ are defined
respectively as
\bear
Q^{I(1)}_\mu(r) &=&
-\frac{\sqrt{mq_I}H_I(r)}{rH^{1/2}(r)}P_\mu^\nu u^\alpha\partial_\alpha
u_\nu
+\frac{1}{2}C^{IJK}\frac{\sqrt{mq_J}\sqrt{mq_K}}{(r^2+q_J)(r^2+q_K)}
\epsilon^{\nu\rho\sigma\mu}u_\nu\partial_\rho u_\sigma \cr &-&
\frac{1}{2r^3H^{1/2}(r)\sqrt{mq_I}}P_\mu^\nu\Big(m(r^2-q_I)\partial_\nu
q_I + q_I(r^2+q_I)\partial_\nu m\Big)\quad,\nonumber\\
P_\mu^{(1)}(r) &=&
\frac{1}{4r^3H(r)}P_\mu^\nu\bigg(H(r)\sum_{I=1}^3\frac{\partial_\nu
q_I}{H_I^2(r)} - \sum_{I=1}^3H_I(r)\sum_{J=1}^3\partial_\nu q_J +
\sum_{I=1}^3H_I(r)\partial_\nu q_I\bigg).
\eear
The $C_\mu$ is the covariantized form of $C_i$ determined before.

We next obtain the first order corrections to the $U(1)^3$ currents, which
haven't been computed before in the literature, and would be the first non-trivial results in this work.
The standard AdS/CFT formula
\be
J^\mu_I =
\lim_{\rho\rightarrow\infty}\frac{\rho^2}{8\pi G_5}\eta^{\mu\nu}A^I_\nu(\rho)\quad,
\ee
works fine here, and it is easy to get the covariantized first order correction
\be
J^{I(1)}_{\mu} = {1\over 16\pi G_5}\left(Q_\mu^{I(1)}(r_H)-{2 \sqrt{m q_I}\over r_H^2 H^{1\over 3}(r_H)} C_\mu\right)\quad,
\ee
where $Q_\mu^{I(1)}(r)$ is defined in the above. Using the covariant version of the conservation
(\ref{constr}),
\be
P^\nu_\mu\partial_\nu m = -4m P^\nu_\mu \left(u^\alpha\partial_\alpha u_\nu\right)\quad,
\ee
and the expression for the density $\rho_I$ in (\ref{0thcurrent}), one can rewrite the result in a more
suggestive form, up to first order in derivatives
\be
J^\mu_{I}= \rho_I u^\mu  -{\cal D}_IP^{\mu\nu} D_\nu \rho_I +\zeta_I
\epsilon^{\nu\rho\sigma\mu}u_\nu\partial_\rho u_\sigma+\cdots \quad,
\ee
where the diffusion coefficient ${\cal D}_I$ is given by
\be
{\cal D}_I = {(r_H^2-q_I)\over 2 r_H^3 H^{1\over 2}(r_H)}\quad,
\ee
and the parity-violating coefficient $\zeta_I$ (originated from the 5D Chern-Simons term) is
\bear
\zeta_I &=& {1\over 32 \pi G_5}\Bigg( C_{IJK} {\sqrt{m q_J} \sqrt{m q_K}
\over (r_H^2+q_J)(r_H^2+q_K)} \nonumber\\
&-& {\sqrt{m q_I}\over 3m} C_{JKL} {\sqrt{m q_J}\sqrt{m q_K} \sqrt{m q_L}
\over (r_H^2+q_J)(r_H^2+q_K)(r_H^2+q_L)}\Bigg) \quad.
\eear
We finish this section by presenting the full covariant form of the gauge potential up to
first order in derivatives;
\bear A^I &=&
\Bigg(\frac{\sqrt{mq_I}}{r^2+q_I}u_\mu +
\int_\infty^rdr'
\frac{H(r)}{r'f(r')H^2_I(r')}\left(Q_\mu^I(r')-Q_\mu^I(r_H)+{2\sqrt{m q_I}\over r_H^2 H^{1\over 3}(r_H)} C_\mu\right) \\ &-&
2\sqrt{mq_I}\int_\infty^rdr'\frac{1}{(r')^3H_I^2(r')}
\int_\infty^{r'}dr''\frac{H(r'')}{r''f^2(r'')}\left(
\int_{r_H}^{r''}dr'''I_\mu(r''')-{r_H f'(r_H)\over H^{1\over 3}(r_H)}C_\mu \right)\Bigg)dx^\mu \quad,\nonumber\eear
where appropriate functions are defined previously.

\section{Hydrodynamics with $SU(2)$ in arbitrary dimensions}

Our next subject is to consider non-Abelian symmetry dynamics
in hot hydrodynamic plasmas, motivated by the iso-spin $SU(2)_I$ dynamics in the QCD plasma.
Although one can embed our bosonic action into the well-defined $AdS_4/CFT_3$ set-up
of the Tri-Sasakian compactification of M-theory to $AdS_4$,
we perform our analysis in arbitrary dimensions envisioning that any system with non-Abelian
symmetry would be described, at least approximately, by our model. In fact, the action we study
has the simplest form one can imagine with gravity and gauge fields in $AdS$.
However, in general dimensions other than $n=4$ the connection to real QCD will no longer be our main motivation.

We will consider $(n+1)$-dimensional gravity corresponding to $n$-dimensional CFT.
We restrict ourselves to the cases of $n\ge 3$ only, as the $n=2$ case seems peculiar in our results.
Our action contains
gravity with $SU(2)$ gauge fields\footnote{One can simply substitute $\epsilon^{abc}$
below with any structure constants of a Lie algebra to have the general results for arbitrary Lie groups.}
\be
{\cal L}={1\over 16\pi G_{n+1}}\left( R+n(n-1) -F^a_{MN}F^{aMN}\right)\quad,
\ee
with
\be
F_{MN}^a = \partial_M A^a_N- \partial_N A^a_M+\epsilon^{abc}A^b_M A^c_N\quad.
\ee
Note that we are allowed to choose
and have chosen a specific normalization for the cosmological constant for simplicity,
while the normalization of the gauge fields in the above corresponds to a definite value of
coupling constant. In $n=3$, this is dictated by the supersymmetry of $N=3$ gauged supergravity \cite{Freedman:1976aw},
and we simply extend it to any dimensions. One can easily recover the gauge coupling constant dependence
in our results below, if needed.
The equations of motion are
\bear
&&R_{MN}+\left(n+{1\over n-1} F^a_{PQ}F^{aPQ}\right)g_{MN}-2 F^a_{PM}{ F^{aP}}_N =0\quad,\nonumber\\
&&\nabla_M {F^{aM}}_N +\epsilon^{abc}A^b_M {F^{cM}}_N =0\quad,
\eear
where $\nabla_M$ is the covariant derivative with metric Christofel connections.

A charged black-brane solution in a general boosted frame is
\bear
ds^2&=& -r^2 V(r)u_\mu u_\nu dx^\mu dx^\nu -2 u_\mu dx^\mu dr + r^2\left(\eta_{\mu\nu}+u_\mu u_\nu\right)
dx^\mu dx^\nu\quad,\nonumber\\
A^a&=& \sqrt{n-1\over 2(n-2)}{q^a\over r^{n-2}} u_\mu dx^\mu\quad,
\eear
with
\be
V(r)=1-{m\over r^n}+{q^a q^a \over r^{2n-2}}\quad.
\ee
where it is simply obtained by embedding the $U(1)$ Reisner-Nordstrom black-brane into
a Cartan direction inside $SU(2)$ which is specified by $q^a$ ($a=1,2,3$).
As we are going to consider slow variations of $q^a$ over the CFT spacetime $x^\mu$,
the Cartan $U(1)$ will correspondingly vary point-by-point, and the non-Abelian nature
will manifest itself when these variations are not parallel to $q^a$ locally.

Let us consider slowly varying parameters $u_\mu$, $m$, and $q^a$ up to first order, and
we work in the frame where $u_\mu=(-1,0,\cdots,0)$ at the position $x^\mu=0$.
Then at first order in derivatives, we have
\bear
u_\mu&=&(-1, x^\mu\partial_\mu u_i)\nonumber\\
m&=& m^{(0)}+x^\mu\partial_\mu m\nonumber\\
q^a&=& q^{a(0)}+x^\mu\partial_\mu q^a\quad,
\eear
and the above black-brane solution will no longer be a solution with these varying parameters.
To be a solution, we have to add  corrections $g^{(1)}_{MN}$ and $A^{a(1)}_M$
to the zero'th order solution with varying parameters,
which should be chosen to satisfy the equations of motion.
Our gauge choice is as before;
\be
g^{(1)}_{rr}=0\quad,\quad g^{(1)}_{r\mu}\sim u_\mu\quad,\quad A^{a(1)}_r=0\quad,
\quad \sum_{i=1}^{n-1}g^{(1)}_{ii}=0\quad.
\ee
The resulting metric and the gauge fields at first order are
\bear
ds^2&=&-r^2V^{(0)}(r)dt^2+2dtdr+r^2 (dx^i)^2\nonumber\\
 &+&\left[x^\mu\left({\left(\partial_\mu m\right)\over r^{n-2}}-{2q^a\left(\partial_\mu q^a\right)\over
 r^{2n-4}}\right)+g^{(1)}_{tt}(r)\right] dt^2+2 g^{(1)}_{tr}(r) dtdr\nonumber\\
 &+&2\left[x^\mu\left(\partial_\mu u_i\right) r^2
 \left(V^{(0)}(r)-1\right)+g^{(1)}_{ti}(r)\right]dtdx^i +2\left[-x^\mu\left(\partial_\mu u_i\right)\right]drdx^i
 \nonumber\\ &+& g^{(1)}_{ij}(r)dx^i dx^j\quad,\\
 A^a&=&-\sqrt{n-1\over 2(n-2)}{q^{a(0)}\over r^{n-2}}dt\nonumber\\
 &+&\left[-\sqrt{n-1\over 2(n-2)}{1\over r^{n-2}}x^\mu\left(\partial_\mu q^a\right)
 +A^{a(1)}_t(r)\right]dt\nonumber\\
 &+&\left[\sqrt{n-1\over 2(n-2)}{q^{a(0)}\over r^{n-2}}x^\mu\left(\partial_\mu u_i\right)+A^{a(1)}_i(r)\right]dx^i
\quad.
\eear

After a lengthy calculation, we get the following equations for the first order corrections
$g^{(1)}_{MN}$ and $A^{a(1)}_M$. From the $tt$-part of Einstein equation,
\bear
&&-{1\over 2}{V(r)\over r^{n-3}}\partial_r\left(r^{n-1} \partial_r g^{(1)}_{tt}\right)-{1\over 2}
r^2 V(r)\partial_r\left(r^2 V(r)\right)\left(\partial_r g^{(1)}_{tr}\right)
-2n \left(r^2 V(r)\right)  g^{(1)}_{tr}\nonumber\\
&&-4(n-2)\sqrt{n-2\over 2(n-1)}{q^a\over r^{n-3}}V(r) \left(\partial_r A^{a(1)}_t\right)\nonumber\\
&&=-{1\over 2}\partial_r\left(r^2 V(r)\right)\left(\partial_i u_i\right)
-{(n-1)\over 2r}\left({\left(\partial_t m\right)\over r^{n-2}}-{2 q^a\left(\partial_t q^a\right)\over r^{2n-4}}\right)
\quad,\label{tt}
\eear
from the $tr$-part,
\bear
&&{1\over 2} {1\over r^{n-1}}\partial_r\left(r^{n-1}\partial_r  g^{(1)}_{tt}\right)
+{1\over 2}\partial_r\left(r^2 V(r)\right)\left(\partial_r g^{(1)}_{tr}\right)
+2n  g^{(1)}_{tr}\nonumber\\
&&+4(n-2)\sqrt{n-2\over 2(n-1)}{q^a\over r^{n-1}}\left(\partial_r A^{a(1)}_t\right)=
{\left(\partial_i u_i\right)\over r}\quad,\label{tr}
\eear
from $rr$-part,
\be
{(n-1)\over r}\left(\partial_r  g^{(1)}_{tr}\right) =0\quad,\label{rr}
\ee
and from $\sum_{i=1}^{n-1}(ii)$-part,
\bear
{(n-1)\over r^{n-3}}\partial_r\left(r^{n-2}  g^{(1)}_{tt}\right)
+(n-1)r^3 V(r)\left(\partial_r  g^{(1)}_{tr}\right) +2n(n-1) r^2  g^{(1)}_{tr}\nonumber\\
-4\sqrt{(n-1)(n-2)\over 2}{q^a\over r^{n-3}}\left(\partial_r A^{a(1)}_t\right)=
2(n-1)r\left(\partial_i u_i\right)\quad.\label{ii}
\eear

From the $t$-part of Maxwell equation, we have
\bear
{V(r)\over r^{n-3}}\partial_r\left(r^{n-1}\partial_r A^{a(1)}_t\right)
&-&\sqrt{(n-1)(n-2)\over 2}{q^a\over r^{n-3}}V(r)\left(\partial_r  g^{(1)}_{tr}\right)
-\sqrt{n-1\over 2(n-2)}\epsilon^{abc}{q^b\over r^{2n-4}}\partial_r\left(r^{n-2} A^{c(1)}_t\right)\nonumber\\
&=& -\sqrt{(n-1)(n-2)\over 2}{1\over r^{n-1}}\left(\left(\partial_t q^a\right)+q^a\left(\partial_i u_i\right)\right)\quad,
\label{t}
\eear
and from the $r$-part,
\be
-{1\over r^{n-1}}\partial_r\left(r^{n-1}\partial_r A^{a(1)}_t\right)
+\sqrt{(n-1)(n-2)\over 2}{q^a\over r^{n-1}}\left(\partial_r  g^{(1)}_{tr}\right)=0\quad.\label{r}
\ee
The above six equations are scalar modes under $SO(n-1)$ spatial rotations.

Vector modes equations of $SO(n-1)$ are the following. From the $ti$-part of Einstein equation,
\bear
-{1\over 2}{V(r)\over r^{n-3}}\partial_r\left(r^{n+1}\partial_r\left( g^{(1)}_{ti}\over r^2\right)\right)
-2\sqrt{(n-1)(n-2)\over 2}{q^a\over r^{n-3}} V(r)\left(\partial_r A^{a(1)}_i\right)\nonumber\\
={1\over 2} {\left(\partial_i m\right)\over r^{n-1}}+
\left({(n-1)\over 2} r V(r) +{n\over 2}{m\over r^{n-1}}\right)\left(
\partial_t u_i\right)\quad,\label{ti}
\eear
from the $ri$-part,
\be
{1\over 2}{1\over r^{n-1}}\partial_r\left(r^{n+1}\partial_r\left( g^{(1)}_{ti}\over r^2\right)\right)
+2\sqrt{(n-1)(n-2)\over 2}{q^a\over r^{n-1}}\left(\partial_r A^{a(1)}_i\right)
=-{(n-1)\over 2r}\left(\partial_t u_i\right)\quad,\label{ri}
\ee
and from the $i$-components of Maxwell equation, we have
\bear
{1\over r^{n-3}}\partial_r\left(r^{n-1} V(r)\partial_r A^{a(1)}_i\right)
&+&\sqrt{(n-1)(n-2)\over 2}{q^a\over r^{n-3}}\partial_r\left(g^{(1)}_{ti}\over r^2 \right)
-2\sqrt{n-1\over 2(n-2)}\epsilon^{abc}{q^b\over r^{n-{5\over 2}}}\partial_r\left(A^{c(1)}_i\over r^{1\over 2}
\right)\nonumber\\
&=&\sqrt{n-1\over 2(n-2)}{1\over r^{n-1}}\left(\left(\partial_i q^a\right)+q^a\left(\partial_t u_i\right)\right).
\label{i}
\eear
Finally, the tensor mode, that is, traceless $ij$-components of Einstein equation is
\be
-{1\over 2}{1\over r^{n-3}}\partial_r\left(r^{n+1} V(r)\partial_r\left( g^{(1)}_{ij}\over r^2\right)\right)
={(n-1)\over 2}r\left(\left(\partial_i u_j\right)+\left(\partial_j u_i\right)-{2\delta_{ij}\over n-1}\left(
\partial_k u_k\right)\right)\quad.\label{ij}
\ee
We now present the complete solution of the above equations.

\subsection{The solution}

We first solve the scalar mode equations. The constraint equations will be discussed after that.
From (\ref{rr}), one finds that $g^{(1)}_{tr}=C$ with some constant $C$. To understand its meaning,
note that this $g^{(1)}_{tr}=C$ will affect only (\ref{tt}), (\ref{tr}), (\ref{ii}), and
it is easy to check that it can be compensated by turning on
\be
g^{(1)}_{tt}=-2C r^2\quad,
\ee
that is, $C$ corresponds to the above homogeneous solution of the problem. As the above $g^{(1)}_{tt}$
is a non-normalizable perturbation to the boundary CFT metric (look at the $r^2$ factor in front),
we see that $C$ in fact corresponds to a non-normalizable homogeneous solution of the problem,
and we set it zero.
Then equation (\ref{r}) is integrated to give us
\be
A^{a(1)}_t = {C^{a}\over r^{n-2} }+ C^{a'}\quad,
\ee
where $C^{a'}=0$ is again a non-normalizable mode, and the meaning of $C^a$ can be
easily understood by looking back the zero'th order profile of the gauge fields
\be
A^{a(0)}_t \sim {q^a \over r^{n-2}}\quad,
\ee
that is, $C^a$ is simply mapped to a redefinition of the charges $q^a$, so that we can also set it zero.
Then one can easily integrate (\ref{ii}) to have
\be
g^{(1)}_{tt}={2\over(n-1)} r \left(\partial_i u_i\right) +{C'\over r^{n-2}}\quad,
\ee
with an integration constant $C'$. However, recall that the zero'th order $g^{(0)}_{tt}$ is
\be
g^{(0)}_{tt} = -r^2 V(r) = -r^2 + {m\over r^{n-2}} -{q^a q^a \over r^{2n-4}}\quad,
\ee
so that $C'$ is simply a redefinition of the energy density $m$. In summary, the only non-vanishing
scalar mode in the solution is
\be
g^{(1)}_{tt}={2\over (n-1)} r \left(\partial_i u_i\right) \quad.
\ee

One then finds the following three kinds of constraints equations,
\bear
0&=& g^{rt} E_{tt}+ g^{rr} E_{rt}= {(n-1)\over 2 r^{n-1}}\left( \left(\partial_t m\right)+{n\over (n-1)} m\left(\partial_i u_i\right)\right)
-{(n-1) q^a\over r^{2n-3}}\left(\left(\partial_t q^a\right)+q^a \left(\partial_i u_i\right)\right)\quad,
\nonumber\\
0&=& g^{rt} M_t^a + g^{rr} M_r^a=\sqrt{(n-1)(n-2)\over 2} {1\over r^{n-1}}
\left(\left(\partial_t q^a\right)+q^a\left(\partial_i u_i\right)\right)\quad,\nonumber\\
0&=& g^{rt} E_{ti} + g^{rr} E_{ri} = -{1\over 2 r^{n-1}}\left(\left(\partial_i m\right) + nm\left(
\partial_t u_i\right)\right)\quad,
\eear
which results in
\be
\left(\partial_t m\right)= -{n\over (n-1)} m \left(\partial_i u_i\right)\quad,\quad
\left(\partial_i m\right)= -n m \left(\partial_t u_i\right)\quad,\quad
\left(\partial_t q^a\right)=- q^a\left(\partial_i u_i\right)\,.
\ee
As before, they are the conservation laws for the zero'th order energy-momentum tensor
and the $SU(2)$ currents;\footnote{We will obtain them more rigorously in the next section.}
\bear
T^{\mu\nu(0)}&=& {m\over 16\pi G_{n+1}}\left(\eta^{\mu\nu}+n u^\mu u^\nu\right)
\equiv p\left(\eta^{\mu\nu}+n u^\mu u^\nu\right)\quad,\nonumber \\
J^{\mu a(0)} &=& {1\over 4\pi G_{n+1}}\sqrt{(n-1)(n-2)\over 2}q^a u^\mu\equiv \rho^a u^\mu \label{0thdensity}
  \quad.
\eear

We next solve for the vector modes, Eqs.(\ref{ri}) and (\ref{i}),
which will turn out to have an important new
ingredient due to the non-Abelian nature. An inspection shows that the only chance to see non-Abelian
nature at this order is when $\left(\partial_i q^a\right)$ is not parallel to $q^a$.
Define
\be
Q^a_i \equiv \epsilon^{abc} q^b \left(\partial_i q^c\right)\quad,
\ee
then the following three $SU(2)$ vectors in the Lie algebra of $SU(2)$ form
a normal basis;
\be
\Big\{ q^a, Q^a_i, \epsilon^{abc} q^b Q^c_i\Big\}\quad,
\ee
and especially one can expand $\left(\partial_i q^a\right)$ in terms of them as
\be
\left(\partial_i q^a\right)= \left(\vec q \cdot \left(\partial_i \vec q\right) \over \vec q\cdot \vec q
\right) q^a +\left(-{1\over \vec q\cdot\vec q }\right) \epsilon^{abc} q^b Q^c_i\quad,
\ee
where $\vec P\cdot \vec Q\equiv P^a Q^a$. We then have to expand our $A^{a(1)}_i$ in terms of them as
\be
A^{a(1)}_i = A^{(1)}_i(r) q^a + f^{(1)}(r) Q_i^a +g^{(1)}(r)\epsilon^{abc} q^b Q_i^c\quad,
\ee
with radial functions $ A^{(1)}_i$, $f^{(1)}$, and $g^{(1)}$ to be determined.
Observe that we have dropped the index $i$ for  $f^{(1)}$ and $g^{(1)}$ as we will see that
they are independent of $i$\footnote{This should be clear from the spatial $SO(3)$ symmetry because
$Q^a_i$ and $\epsilon^{abc}q^b Q_i^c$ are already $SO(3)$ vectors.}.
Plugging this expansion into our equations (\ref{ri}) and (\ref{i}), one obtains
the following two equations for $A^{(1)}_i$ and $g_{ti}^{(1)}$,
\bear
{1\over 2}{1\over r^{n-1}}\partial_r\left(r^{n+1} \partial_r\left(g^{(1)}_{ti}\over r^2\right)\right)
+2\sqrt{(n-1)(n-2)\over 2}{\vec q\cdot \vec q \over r^{n-1}} \left(\partial_r A^{(1)}_i\right)
= -{(n-1)\over 2r} \left(\partial_t u_i\right),\nonumber\\\label{rinew}
\eear
\bear
{1\over r^{n-3}}\partial_r\left(r^{n-1} V(r) \partial_r A^{(1)}_i \right)
+\sqrt{(n-1)(n-2)\over 2}{1\over r^{n-3}}\partial_r\left(g^{(1)}_{ti}\over r^2\right)\nonumber\\
=\sqrt{n-1\over 2(n-2)} {1\over r^{n-1}}\left({
\vec q \cdot \left(\partial_i \vec q\right) \over \vec q\cdot \vec q }+\left(\partial_t u_i\right)\right)\quad,\label{inew}
\eear
and the following coupled equations for $f^{(1)}$ and $g^{(1)}$,
\bear
{1\over r^{n-3}}\partial_r\left(r^{n-1} V(r)\partial_r f^{(1)}\right)
+2\sqrt{n-1\over 2(n-2)}{\vec q\cdot\vec q \over r^{n-{5\over 2}}}\partial_r \left(g^{(1)}\over r^{1\over 2}\right)=0\,,\label{fnew}
\eear
\bear
{1\over r^{n-3}}\partial_r\left(r^{n-1} V(r)\partial_r g^{(1)}\right)
-2\sqrt{n-1\over 2(n-2)}{1  \over r^{n-{5\over 2}}}\partial_r\left( f^{(1)}\over r^{1\over 2}\right)=
-\sqrt{n-1\over 2(n-2)} {1\over r^{n-1}} {1\over \vec q\cdot\vec q}\,.\label{gnew}
\eear

It is straightforward to solve (\ref{rinew}) and (\ref{inew}) as is done in the STU model, and we simply
present the result;
\bear
g^{(1)}_{ti}&=& r^2V(r)\int_\infty^r dr'\,
{1\over (r')^{n+1} (V(r'))^2}\left(\int_{r_H}^{r'} dr''\, I(r'')-r_H^{n-1} V'(r_H) C_i\right)\quad,\\
A^{(1)}_i&=& \int_\infty^r dr'\,{1\over (r')^{n-1} V(r')}\left(Q_i(r')-Q_i(r_H)
-\sqrt{(n-1)(n-2)\over 2}\left({g^{(1)}_{ti}(r')\over (r')^2}-{C_i\over r_H^2}\right)\right)\quad,\nonumber
\eear
where
\bear
Q_i(r)&=&-\sqrt{n-1\over 2(n-2)}{1\over r}\left({
\vec q \cdot \left(\partial_i \vec q\right) \over \vec q\cdot \vec q }+\left(\partial_t u_i\right)\right)\quad,\nonumber\\
I(r)&=&-4\sqrt{(n-1)(n-2)\over 2}{\vec q\cdot\vec q \over r^{n-1}}\left(Q_i(r)-Q_i(r_H)+\sqrt{(n-1)(n-2)\over 2}{C_i\over r_H^2}\right)\nonumber\\
&-&(n-1)r^{n-2}V(r)\left(\partial_t u_i\right)\,.\nonumber
\eear
Again, the integration constant $C_i$ is fixed by the Landau frame constraint, and an explicit computation gives us
\bear
C_i&=& {r_H^2\over n m}\Bigg(-{2\over (n-2) r_H^{n-1}}\left(\vec q\cdot \partial_i \vec q\right)
+\left(n r_H^{n-1}+{(n^2-4n+2) \vec q\cdot\vec q \over (n-2) r_H^{n-1}}\right)\left(\partial_t u_i\right)\Bigg).\nonumber
\eear
However, we are unable to integrate the equations (\ref{fnew}) and (\ref{gnew})
to solve $f^{(1)}$ and $g^{(1)}$; we will instead
comment on a possible numerical approach. Defining $F^{(+)(1)}\equiv f^{(1)}+i|\vec q|g^{(1)}$,
the equations (\ref{fnew}) and (\ref{gnew}) become a single complex equation
\bear
{1\over r^{n-3}}\partial_r\left(r^{n-1} V(r)\partial_r F^{(+)(1)}\right)
-2i|\vec q|\sqrt{n-1\over 2(n-2)}{1  \over r^{n-{5\over 2}}}\partial_r\left( F^{(+)(1)}\over r^{1\over 2}\right)=
-i|\vec q|\sqrt{n-1\over 2(n-2)} {1\over r^{n-1}} {1\over \vec q\cdot\vec q}\,.\nonumber
\eear
The equation has one trivial solution,
\be
F^{(+)(1)} = -{1\over \vec q\cdot \vec q}\quad,
\ee
which is non-normalizable. However, having this solution is of great help in finding
the unique normalizable solution regular at the horizon numerically; one only solves the {\it homogeneous}
equation without the source term in the right-hand side, and then add the above trivial solution
to have a normalizable solution.
Considering the limit of the {\it homogeneous} equation to the horizon where $V(r_H)=0$, one gets
the relation
\be
{\left(\partial_r F^{(+)(1)}\right)\over F^{(+)(1)}}\Bigg|_{r=r_H}
={-i|\vec q|\sqrt{n-1}\over
\sqrt{2(n-2)}r_H^{n+1} V'(r_H) - 2i\sqrt{n-1} |\vec q|r_H}\quad,
\ee
for a regular homogeneous solution. Putting $F^{(+)(1)}(r_H)=1$, the above completely specifies
the boundary condition at the horizon, and one can numerically solve the differential equation uniquely,
that is, regular homogeneous solution normalized as $F^{(+)(1)}(r_H)=1$ is unique. Let's call
this homogeneous solution $F^{(+)(1)}_0(r)$. In general, its large $r$ asymptotic will
give us a finite non-zero constant $F^{(+)(1)}_0(\infty)\ne 0$, and one constructs the full solution
of our original problem simply as
\be
F^{(+)(1)}(r)= -{1\over \vec q\cdot\vec q}\left(1-{F^{(+)(1)}_0(r)\over F^{(+)(1)}_0(\infty)}\right)\quad.
\ee

Finally, the tensor mode equation (\ref{ij}) is easily integrated to give us
\be
g^{(1)}_{ij}=-2\sigma_{ij} r^2 \int^r_\infty dr'\,{1\over (r')^{n+1} V(r')}\left(
(r')^{n-1} - (r_H)^{n-1}\right)\quad,
\ee
with
\be
\sigma_{ij}={1\over 2}\left(\left(\partial_i u_j\right)+\left(\partial_j u_i\right)-{2\delta_{ij}\over n-1}\left(
\partial_k u_k\right)\right)\quad.
\ee

\subsection{The first order transport coefficients}

Based on the results of the previous section, we can write the covariant form of the metric
and the gauge field up to first order in derivative expansion. The
metric looks as
\bear ds^2 &=&
- r^2V(r)u_\mu u_\nu dx^\mu dx^\nu -2u_\mu dx^\mu dr+ r^2 P_{\mu\nu} dx^\mu dx^\nu
 + \frac{2}{n-1}r(\partial_\rho u_\rho) u_\mu u_\nu dx^\mu
dx^\nu \nonumber\\& -& r^2 V(r)u_\mu
\bigg(\int_\infty^rdr'\frac{1}{(r')^{n-1}V^2(r')}\left(\int_{r_H}^{r'}dr''I_\nu(r'')-r_H^{n-1} V'(r_H) C_\mu\right)\bigg)
dx^\mu dx^\nu \cr &-& 2r^2\sigma_{\mu\nu}\bigg(\int_\infty^r
dr'\frac{1}{(r')^{n+1}V(r')}\Big((r')^{n-1}-(r_H)^{n-1}\Big)\bigg)
dx^\mu dx^\nu \quad,
\eear
where the first-order functions
$I^{(1)}_\mu(r)$ and $Q^{(1)}_\mu(r)$ are defined respectively as
\bear I^{(1)}_\mu(r) &=&
-4\sqrt{\frac{(n-1)(n-2)}{2}}\frac{q.q}{r^{n-1}}\Bigg(Q^{(1)}_\mu(r)-Q^{(1)}_\mu(r_H)+
\sqrt{\frac{(n-1)(n-2)}{2}}{C_\mu\over r_H^2}\Bigg)\nonumber\\
&+&(n-1)r^{n-2}V(r)u^\nu\partial_\nu u_\mu \quad,\nonumber\\
Q^{(1)}_\mu(r) &=&
-\sqrt{\frac{n-1}{2(n-2)}}\frac{1}{r}\bigg(\frac{q\cdot
P_\mu^\nu\partial_\nu q}{q\cdot q} - u^\nu\partial_\nu u_\mu\bigg)\label{Qmu}
\quad,
\eear
and we also defined
\be
\sigma^{\mu\nu} = \frac{1}{2}P^{\mu\alpha}P^{\nu\beta}\left(\partial_\alpha u_\beta + \partial_\beta
u_\alpha\right) -\frac{P^{\mu\nu}}{n-1}\left(\partial_\alpha u^\alpha\right) \quad.
\ee
The gauge field is found as
\bear A^a &=&
q^a\Bigg(\sqrt{\frac{n-1}{2(n-2)}}\frac{1}{r^{n-2}}u_\mu
+ \int_\infty^r
dr'\frac{1}{(r')^{n-1}V(r')}\Bigg(Q_\mu^{(1)}(r')-Q_\mu^{(1)}(r_H)\nonumber\\&+&\sqrt{\frac{(n-1)(n-2)}{2}}{C_\mu\over r_H^2}\Bigg)-
\sqrt{\frac{(n-1)(n-2)}{2}}\int_\infty^rdr'\frac{1}{(r')^{n-1}}\nonumber\\
&&\int_\infty^{r'} dr''\frac{1}{(r'')^{n+1}V^2(r'')}
\left(\int_{r_H}^{r''}dr'''I^{(1)}_\mu(r''')-r_H^{n-1} V'(r_H) C_\mu\right) \Bigg)dx^\mu\nonumber\\
&+&\left(f^{(1)}(r)Q_\mu^a +g^{(1)}(r)\epsilon^{abc}q^b Q_\mu^c\right)dx^\mu
\quad,\nonumber\eear where $Q^a_\mu \equiv \epsilon^{abc}q^b P^\nu_\mu(\partial_\nu q^c)$.
Note that the last line is the non-Abelian induced terms.

Again using holographic renormalization, one can easily find the stress
tensor and the $SU(2)$ charge current.
In terms of Fefferman-Graham coordinate expansion,
\be
g_{\mu\nu}(\rho)=\eta_{\mu\nu}+\cdots + {g^{(n)}_{\mu\nu}\over \rho^n}+\cdots\quad,
\ee
the $n$-dimensional CFT stress tensor is given by
\be
T_{\mu\nu}={n\over 16\pi G_{n+1}}g^{(n)}_{\mu\nu}\quad.
\ee
Again, it is straightforward to check that the only non-vanishing first order correction
comes from $g_{ij}^{(1)}$ with
\be
T_{ij}^{(1)} = -2\eta \sigma_{ij}\quad,
\ee
where
\be
\eta= {r_H^{n-1}\over 16\pi G_{n+1}} = {s\over 4\pi}\quad.
\ee
We point out that it is not a trivial fact here and in the STU model
that there is no first order correction to $T^{00}$,
as it results from a non-trivial cancellations in the expansion.
The full covariant expression of the stress tensor is
\be
T^{\mu\nu} = p\left(\eta^{\mu\nu}+n u^\mu u^\nu\right)-2\eta \sigma^{\mu\nu} +\cdots\quad.
\ee
with the pressure $p={m\over 16\pi G_{n+1}}$ representing the zero'th order contribution.

The $SU(2)$ charge current is similarly obtained from the expansion as
\be
J^a_\mu =  {(n-2)\over 4\pi G_{n+1}}\lim_{\rho\to\infty}\rho^{n-2} A^a_\mu(\rho)\quad,
\ee
and it is easy to find the first order correction to be
\bear
J_\mu^{a(1)} = {1\over 4\pi G_{n+1}}\left(Q^{(1)}_\mu (r_H) -  \sqrt{(n-1)(n-2)\over 2}{C_\mu\over r_H^2}\right) q^a
+{(n-2)\over 4\pi G_{n+1}}\left(f^{(n-2)} Q^a_\mu +g^{(n-2)}\epsilon^{abc} q^b Q_\mu^c\right),\nonumber
\eear
where $Q^{(1)}_\mu (r)$ is given before in (\ref{Qmu}),
and $f^{(n-2)}$, $g^{(n-2)}$ are the coefficients of $1/\rho^{n-2}$ in the
expansion of $f^{(1)}(\rho)$ and $g^{(1)}(\rho)$ respectively.
Trading $q^a$ with the density $\rho^a$ defined in (\ref{0thdensity}), we can rewrite
the result in the form
\bear
J^{a(1)}_\mu &=&
-{\cal D} \left({\rho\cdot P^\nu_\mu (\partial_\nu\rho)\over \rho\cdot\rho}-u^\nu\partial_\nu u_\mu\right)
\rho^a+{\cal D}_1 \epsilon^{abc}\rho^b P^\nu_\mu (\partial_\nu \rho^c)\nonumber\\
&+&{\cal D}_2P^\nu_\mu\left(\rho^a (\rho\cdot
\partial_\nu\rho)-(\rho\cdot\rho) (\partial_\nu\rho^a) \right)\quad,
\eear with three diffusion coefficients 
\bear {\cal D} &=& {1\over
(n-2) r_H}\left(1-{2 \vec q\cdot \vec q \over n m r_H^{n-2}}\right)={(n-2)m+2 r_H^n \over n(n-2)m r_H} \quad,\nonumber\\
 {\cal D}_1&=& {8\pi G_{n+1} f^{(n-2)}\over
(n-1)} \quad,\quad {\cal D}_2={2^{11\over 2} \pi^2 G_{n+1}^2
g^{(n-2)}\over (n-1)^{3\over 2}(n-2)^{1\over 2}}\quad. \eear The
${\cal D}$ is essentially the usual diffusion coefficient of Abelian
nature, which agrees with ref.\cite{Starinets:2008fb},
while the other two diffusion coefficients are due to the
non-Abelian properties. Although their precise values can only be
determined by numerical analysis, we hope that the above structure
of non-Abelian current we obtain in derivative expansion may be
important for future applications to the QCD plasma incorporating
non-Abelian symmetry.

\subsection{On Tri-Sasakian compactification of M-theory to $AdS_4$}

We would like to conclude  by a few comments on the realization of our $SU(2)$ theory in a concrete example.
Our 4-dimensional bosonic action ($n=3$) with $SU(2)$ gauge symmetry in the bulk has a
specific $AdS_4/CFT_3$ realization in M-theory; consider $N$ M2 branes sitting at the apex of
an 8-dimensional Hyper-Kahler cone and take a near horizon limit. The superconformal theory one
gets on the M2 branes has $N=3$ or 6-real components of dynamical supersymmetry with $SU(2)_R$ R-symmetry.
The corresponding dual theory on $AdS_4$ will include a consistent truncation to the minimal
$N=3$ $SU(2)$ gauged supergravity in 4-dimensions\cite{Freedman:1976aw}, whose bosonic action is precisely our action in the previous section.
In this case, one has an explicit expression for the Newton's constant $G_4$ in terms
of the number of M2 branes as follows.

The 11-dimensional M-theory supergravity action is
\be
{\cal L}_{11}= {1\over (2\pi)^8 l_p^9} \int d^{11}x\,\sqrt{-g_{11}}\left( R^{(11)}-{1\over 2}|F_4|^2\right)
-{1\over 6(2\pi)^8 l_p^9}\int C_3\wedge F_4\wedge F_4\quad,
\label{maction}
\ee
where $R^{(11)}$ is the Ricci scalar of the 11-dimensional metric, and
\be
|F_4|^2 = {1\over 4 !} F_{MNPQ} F^{MNPQ}\quad.
\ee
The near horizon limit of M2 branes at the tip of a Hyper-Kahler cone takes a form
of $AdS_4\times X_7$ with $X_7$ being a Tri-Sasakian 7-fold which is the unit radius section of the Hyper-Kahler
cone involved. The explicit solution is given as
\bear
ds_{11}^2 &=& R^2\left({1\over 4} ds^2_{AdS_4} + d\Omega^2_{X_7}\right)\nonumber\\
F_4&=& {3\over 8}R^3  \epsilon_4\quad,\label{11d}
\eear
where $d\Omega^2_{X_7}$ is the metric of $X_7$ normalized in such a way that
\be
R_{ab}=6 g_{ab}\quad,
\ee
and $\epsilon_4$ is the volume form of the unit radius $AdS_4$. The constant $R$ is given by the relation
\be
6 R^6 {\rm vol}(X_7) = (2\pi l_p)^6 N\quad.\label{relation}
\ee
Then one can easily obtain the 4-dimensional effective action on $AdS_4$
after compactifying M-theory action (\ref{maction})
on the above 7-dimensional Tri-Sasakian manifold $X_7$.
Note that the $X_7$ metric of $R^2 d\Omega_{X_7}^2$ now has
\be
R_{ab}^{X_7}={6\over R^2} g_{ab}^{X_7}\quad,
\ee
so that $R^{X_7}={42\over R^2}$, which means
\be
R^{(11)} \sim R^{(4)} + {42\over R^2}\quad.
\ee
Also one has
\be
|F_4|^2 = \left(\frac38 R^3\right)^2\left(4\over R^2\right)^4={36\over R^2}\quad.
\ee
Combined with
\be
\int d^{11}x \,\sqrt{-g_{11}}=R^7 {\rm vol}(X_7) \int d^4 x\,\sqrt{-g_4}\quad,
\ee
the effective 4-dimensional action takes a form
\be
{R^7 {\rm vol}(X_7)\over (2\pi)^8 l_p^9}\int d^4 x\,\sqrt{-g_4}\left(R^{(4)} +{24\over R^2}+\cdots \right)
\quad,
\ee
and we identify
\be
{1\over 16\pi G_4}={R^7 {\rm vol}(X_7)\over (2\pi)^8 l_p^9}\quad.
\ee
We then need to put $R=2$ to conform to our convention of cosmological constant
in the previous section\footnote{The quicker way to arrive at this conclusion is to make the $AdS_4$
in (\ref{11d}) to have unit radius, which should be a solution with our cosmological constant convention.},
and
using (\ref{relation}) one finally has
\be
{1\over 16\pi G_4}={2\pi N^{3\over 2}\over 2^2 6^{3\over 2} \left({\rm vol}(X_7)\right)^{1\over 2}}\quad.
\ee

For a class of Tri-Sasakian manifolds that are obtained from Hyper-Kahler quotients,
their normalized volumes ${\rm vol}(X_7)$ are explicitly known\cite{Lee:2006ys,Yee:2006ba}.
One can start from a $(2+r)$-dimensional flat quaternion space and take $U(1)^r$
Hyper-Kahler quotients specified by charges $Q^i_a$ where $i$ runs over $U(1)$ and $a$ runs
over $(2+r)$ quaternions. The resulting 8-dimensional Hyper-Kahler cone will
have a unit radius section as a Tri-Sasakian manifold, whose normalized volume is known by the formula\cite{Yee:2006ba}
\be
 {\rm vol}\left(X_{7}\right)
={2^{r+1} \pi^{4}\over \Gamma(4) {\rm Vol}\left(U(1)^r\right)}\int
\prod_{i=1}^r d\phi^i \,\,\prod_{a=1}^{2+r} {1\over 1+\left(\sum_{i=1}^r Q_a^i
\phi^i\right)^2}\quad.
\ee
In the simplest example of $r=1$ with three charges $Q_i$ ($i=1,2,3$), it becomes
\be
{\rm vol}\left(X_7(Q_1,Q_2,Q_3)\right)={\pi^4\over 3}
{(Q_1Q_2+Q_2Q_3+Q_3Q_1)\over
(Q_1+Q_2)(Q_2+Q_3)(Q_3+Q_1)}\quad,
\ee
which includes the famous $N(1,1)$ as $Q_1=Q_2=Q_3=1$ with
\be
{\rm vol}\left(N(1,1)\right)={\pi^4 \over 8}\quad.
\ee
Although the $N=3$ superconformal theory on M2 branes dual to the gravity background with $N(1,1)$ is
still unknown, there is a proposal in ref.\cite{Jafferis:2008qz} for the case of $Q_1\pm Q_2 \pm Q_3 =0$, inspired
by the theory of BL/ABJM\cite{Bagger:2007jr,Aharony:2008ug}. See also ref.\cite{Lee:2006hw} for $N=2$ cases.

\vskip 1cm \centerline{\large \bf Acknowledgement} \vskip 0.5cm

We would like to thank Mukund Rangamani for many critical discussions and encouragement. We also thank Chris Herzog for
helpful comments.

\appendix
\section{Appendix 1}

A convenient expression for the first order variation of the Ricci tensor we use is
\be
\delta R_{MN}= -\nabla_M C^P_{PN} + \nabla_{P} C^P_{MN}\quad,
\ee
with
\be
C^P_{MN}={1\over 2}g^{PQ}\left(\nabla_M \delta g_{NQ}+\nabla_N \delta g_{MQ} - \nabla_{Q}\delta g_{MN}\right)\quad,
\ee
where $\nabla_M$ is the covariant derivative with respect to the zero'th order metric $g_{MN}$.
The tensor $C^P_{MN}$ is in fact a variation of the metric Christoffel symbol, $C^P_{MN}=\delta \Gamma^P_{MN}$.

\section{Appendix 2}

We denote
\be
A(r)=H^{-{2\over 3}}(r)f(r)\quad,\quad B(r)= H^{-{1\over 6}}(r)\quad,\quad C(r)=r^2 H^{1\over 3}(r)  \quad,
\ee
as before, where the local parameters $m$ and $q_I$ are implicit in the expressions above and below.
The prime in the equations below means the radial derivative ${d\over dr}$.

\bear P_i^{(1)}&=&
{1\over 4r^3 H(r)}\left(H(r)\left(\sum_{I=1}^3 {\partial_i q_I\over H_I^2(r)}\right)
-\left(\sum_{I=1}^3 H_I(r)\right)\left(\sum_{J=1}^3 \partial_i q_J\right)
+\sum_{I=1}^3 \left(H_I(r)\partial_i q_I\right)\right)\,,\nonumber
\eear

\bear
S^{(1)}_{tt} = -{3A'\over 4B C}\left(\partial_t C\right)-A\left({1\over B^2}\left(
\partial_t B\right)\right)'-\left({B'\over B}+{3 C'\over 2C}\right){A\over B^2}
\left(\partial_t B\right)+{3C'\over 4 BC}\left(\partial_t A\right) -{A'\over 2B}\left(\partial_i u_i\right)\,,\nonumber
\eear

\bear
S^{(1)}_{tr}=\left({1\over B}\left(\partial_t B\right)+{3\over 2C}\left(\partial_t C\right)\right)'
+{3C'\over 4C^2}\left(\partial_t C\right)+{1\over 2}\sum_{I=1}^3 {1\over (X^I)^2}\left(\partial_r X^I\right)
\left(\partial_t X^I\right)+
{C'\over 2C}\left(\partial_i u_i\right)\,,\nonumber
\eear

\bear
\sum_i S^{(1)}_{ii}= {3\over B}\left(\partial_t C\right)' +{3C'\over 2 BC}\left(\partial_t C\right)
+{3C'\over B}\left(\partial_i u_i\right)\,,\nonumber
\eear

\bear
S^{1(1)}&=&
-{1\over B}\left({\left(\partial_t X^1\right)\over X^1}+{1\over 2}{\left(\partial_t X^2\right)\over X^2}
\right)'-\left({1\over B}\left({\left(\partial_t X^1\right)\over X^1}+{1\over 2}{\left(\partial_t X^2\right)\over X^2}
\right)\right)'\nonumber\\
&-&{1\over B}\left({B'\over B}+{3C'\over 2C}\right)\left({\left(\partial_t X^1\right)\over X^1}
+{1\over 2}{\left(\partial_t X^2\right)\over X^2}\right)
-{3\over2 BC}\left(\partial_t C\right)
\left({(X^1)' \over X^1}+{1\over 2}{(X^2)'\over X^2}\right)\nonumber\\
&-&{1\over B}\left({(X^1)' \over X^1}+{1\over 2}{(X^2)'\over X^2}\right)\left(\partial_i u_i\right)\,.\nonumber
\eear

 \vfil

\end{document}